\documentclass[journal]{IEEEtran}
\IEEEoverridecommandlockouts
\usepackage{cite}
\usepackage{amsmath,amssymb,amsfonts}
\usepackage{graphicx}
\usepackage{textcomp}
\usepackage{mathtools}
\usepackage{enumitem}
\usepackage{epsfig,subfigure}
\usepackage{multirow}
\usepackage{float}
\usepackage{booktabs}
\usepackage{tikz}
\usepackage[normalem]{ulem}
\usepackage[linesnumbered, ruled, vlined]{algorithm2e}
\usepackage{algpseudocode}
\usepackage{adjustbox}
\usepackage{makecell}
\usepackage{todonotes}
\usepackage{hyperref}
\usepackage{acronym}
\usepackage{xcolor}

\newcommand{\rev}[1]{#1}
\newcommand{\revstart}{}

\newacro{CPS}{cyber-physical systems}
\newacro{MTD}{Moving target defense}
\newacro{RL}{reinforcement learning}
\newacro{ML}{machine learning}
\newacro{MDP}{Markov decision process}
\newacro{FDI}{False data injection}
\newacro{DRL}{deep reinforcement learning}
\newacro{HIL}{hardware-in-the-loop}
\newacro{IoT}{Internet of Things}
\newacro{AGC}{automatic generation control}
\newacro{EMS}{energy management system}
\newacro{PMU}{phasor measurement units}
\newacro{PDC}{phasor data concentrators}
\newacro{WAN}{wide-area networks}
\newacro{MiTM}{man-in-the-middle }
\newacro{DQN}{deep Q-network}
\newacro{SAC}{soft-actor-critic}
\newacro{HDDA}{hybrid deceptive data acquisition}
\newacro{CSER}{Compromised Sensor Elimination Rate }
\newacro{PSA}{Percentage of Stealthy Attacks}
\newacro{PEA}{Percentage of Exposed Attacks }
\newacro{EDR}{Estimation Deviation Rate }
\newacro{ML}{machine learning}
\newacro{MIP}{mixed interger programming}
\newacro{SE}{state estimation}
\newacro{BDD}{bad data detection}
\newacro{WLS}{weighted least square}
\newacro{EKF}{extended kalman filter}

\def\BibTeX{{\rm B\kern-.05em{\sc i\kern-.025em b}\kern-.08em
    T\kern-.1667em\lower.7ex\hbox{E}\kern-.125emX}}
\usepackage{xcolor}
\def\BibTeX{{\rm B\kern-.05em{\sc i\kern-.025em b}\kern-.08em
    T\kern-.1667em\lower.7ex\hbox{E}\kern-.125emX}}

\begin{document}


\title{ROAM: Reinforcement Learning Optimized Adaptive Moving Target Defense for Smart Grid}


\author{{Mohammad Zakaria Haider, \IEEEmembership{Student Member, IEEE}, Prabin~Mali, \IEEEmembership{ Member, IEEE} \\ Muhammad~Nadeem, \IEEEmembership{Member, IEEE}, Sumit~Paudyal,~\IEEEmembership{Senior Member,~IEEE}, \\ Fei~Ding,~\IEEEmembership{Senior Member,~IEEE}, Mohammad Ashiqur Rahman, \IEEEmembership{Senior Member, IEEE}}

\thanks{This work is supported by the Department of Energy (DOE) under Award\# DE-CR0000024. Any opinions, findings, conclusions, or recommendations expressed in this material are those of the authors and do not necessarily reflect the views of the DOE or the U.S. Government.
    
    Mohammad Zakaria Haider is a Ph.D. student in the Department of Electrical and Computer Engineering (ECE) at Florida International University (FIU) and a member of FIU's Analytics for Cyber Defense (ACyD) Lab, Miami, FL, USA (email: mhaid010@fiu.edu ).
    
    Prabin Mali is a Ph.D. student in the Department of Electrical and Computer Engineering (ECE) at Florida International University (FIU) and a member of FIU's Power System Computational Laboratory, Miami, FL, USA (email:pmali004@fiu.edu).
    
    Muhammad Nadeem is a Postdoctoral Associate at Knight Foundation School of Computing and Information Sciences (KFSCIS) at Florida International University and a member of the ACyD Lab, Miami, FL, USA (email: mnadeem@fiu.edu).
    
    Sumit Paudyal is a Professor in the Department of Electrical and Computer Engineering (ECE) at Florida International University and the Director of the Power System Computational Laboratory, Miami, FL, USA (email: spaudyal@fiu.edu).
    
    Fei Ding is with the Power Systems Engineering Center, National Laboratories of the Rockies, Golden, CO, USA (email: fei.ding@nlr.gov).

    Mohammad Ashiqur Rahman is an Associate Professor of the KFSCIS at Florida International University, and the Director of the ACyD Lab, Miami, FL, USA (email: marahman@fiu.edu). 
   	}
}

\maketitle

\begin{abstract}
Smart grids rely on distributed phasor measurement units (PMUs) and wide-area communication networks whose data-acquisition pipelines are vulnerable to stealthy false data injection (FDI) attacks that evade conventional bad-data detection (BDD). Moving target defense (MTD) counters this exposure by continuously remapping sensor identities, but existing schemes apply static or pre-scheduled randomization and are each constrained by a single failure mode: seed-based methods are fast yet fragile to key compromise, while machine learning (ML) based prediction methods are resilient yet incur prohibitive latency. Neither adapts its randomization to live threat evidence. This paper presents ROAM, a deep reinforcement learning (DRL)-driven MTD framework in which a deep Q-network (DQN) agent continuously re-optimizes the sensor-to-ID mapping at the aggregation layer. At each decision step, the agent \rev{selects the deception coverage, flags the specific sensors to remap, and sets the split of the remapped sensors between seed-based and ML-prediction-based recovery}, without modifying PMU hardware and while preserving IEEE C37.118 compatibility. \rev{On held-out BDD-compliant FDI scenarios across the IEEE 14-, 57-, and 118-bus systems, ROAM keeps the percentage of stealthy attacks below 3\% while sustaining detection rates above 93\%. Against an adaptive soft actor-critic (SAC) attacker that observes the defender's configuration, ROAM keeps stealthy attacks below 2.6\% with detection above 92\%.} Hardware-in-the-loop testing on an OPAL-RT platform confirms real-time feasibility, with sub-18 ms \rev{average} shuffle execution and \rev{an average reshuffle blind window of 2.6\% of the 3-s decision cycle}.

\end{abstract}


\section{Introduction}

 \IEEEPARstart{M}{odern} smart grids rely on distributed phasor measurement units (PMUs) and wide-area communication networks for real-time monitoring, \ac{SE}, and \ac{AGC} within an \ac{EMS}. Synchrophasor technology has sharpened situational awareness, but the distributed, networked nature of PMU-based systems exposes data-acquisition pipelines to cyber threats that target measurement integrity and grid stability~\cite{Sadreazami18SIPN, mohammadi2018attack, Pasqualetti, Swain24Access}. \ac{FDI} attacks are especially dangerous: by exploiting knowledge of the system topology and measurement correlations, an adversary can craft perturbations that bypass conventional \ac{BDD} while corrupting \ac{AGC} for frequency stability~\cite{Jafari22IAS, Jafari23TSG, Ran24TII}. Real-world incidents, from Stuxnet~\cite{stuxnet} and Dragonfly~\cite{Dragonfly} to the 2017 Triton/Trisis compromise of safety instrumented systems~~\cite{triton18}, confirm the threat is concrete: the energy sector accounted for 10.7\% of all reported cyberattacks in 2022~\cite{jenifer}. These attacks exploit communication protocols common to energy management systems (EMS), including C37.118~\cite{Knesek21SmartGridComm, Vahidi23ICST}, DNP3~\cite{Wlazlo21IET}, and Modbus~\cite{HUITSING200837, Chen15CQR}, and can be launched through a man-in-the-middle (MITM) attack compromising phasor data concentrators (PDCs) connected to spoofed PMUs~\cite{Saini22NPSC, He21EI2}.

Conventional security measures struggle against such adversaries because their static configurations present a fixed attack surface that a persistent attacker can learn and exploit. \ac{MTD} addresses this by continuously altering system configurations, physical parameters, measurement pathways, or sensor-to-ID mappings so that any intercepted topology or state becomes obsolete before it can be weaponized~\cite{ZhangTSG23, LakshminarayanaTSG21, MohammadpourfardTII25, PagnottaTIFS23, Liu21SmartGridComm, Liu18PESGM}. For power systems, deception-based data-acquisition frameworks such as DDAF~\cite{ddaf21} and iDDAF~\cite{iddaf21} randomize sensor IDs at the sensor layer, converting stealthy FDI vectors into random injections that can be detected and localized by outlier analysis. Parameter-perturbation methods instead alter grid impedance or topology to invalidate the attacker's system model~\cite{MohammadpourfardTII25, ZhangTSG23}, and adversarial DRL has been applied to automate MTD reconfiguration in network and CPS settings~\cite{Eghtesad20Verlag, PagnottaTIFS23, Cao24IoT, Kim22TNSM, Chen24NOMS}.

Beyond heuristic and \ac{RL}-based approaches, a body of work frames the MTD problem as a Stackelberg or Nash game between defender and attacker: Lakshminarayana et al.~\cite{LakshminarayanaTSG21} derive optimal topology-perturbation strategies under a bilevel formulation, complementary graph-theoretic approaches \cite{graphmtd1, graphmtd2} characterize MTD effectiveness via the spectral properties of the measurement Jacobian, and physics-informed anomaly detection methods \cite{pinn_anomaly} embed power-flow physics into \ac{BDD} to reduce false positives under non-Gaussian noise. This paper proposes ROAM, a \uline{R}einforcement Learning \uline{O}ptimized \uline{A}daptive \uline{M}oving Target Defense framework that differs from these approaches by learning the remapping policy \rev{directly} from system feedback, eliminating the need for explicit equilibrium computation or hand-crafted graph features, while incorporating real-time threat evidence that static bilevel formulations cannot accommodate.

Despite this progress, existing sensor-remapping MTD schemes exhibit complementary failure modes that have not been addressed jointly. Seed-based methods use cryptographically secure pseudorandom generators to deterministically alter sensor IDs, achieving sub-2 ms recovery with low overhead; however, they depend on pre-shared keys and synchronized state, leaving them vulnerable to key compromise and desynchronization~\cite{ddaf21, iddaf21}. On the other hand, \rev{ML-based} remapping prediction removes the shared secret by learning mappings from historical data and threat patterns, enabling adaptation to evolving attacks; however, its combinatorial matching step increases latency by roughly 60$\times$ and can degrade accuracy in resource-constrained settings~\cite{tan2022wf, hamlet2016dependency}. More fundamentally, both families commit to a fixed or pre-scheduled randomization policy and therefore leave the operative questions unanswered: how many sensors to remap, which sensors to prioritize, and when to escalate from lightweight shuffling to heavier prediction-based recovery. These decisions should respond to evidence of live threats, yet no current scheme \rev{adapts its remapping strategy} to real-time threat assessment~\cite{Huang19Verlag, Rahman14ACM}.

Our proposed ROAM closes this loop by treating the sensor-to-ID mapping as a continuously re-optimized decision variable. A \ac{DQN} agent observes the filtered system state and \ac{BDD} feedback, then \rev{selects, at each decision epoch, the deception coverage, flags which sensors are remapped, and sets the fraction of remapped sensors recovered through the seed-based versus the ML-prediction-based pathway.} The framework operates entirely at the PDC aggregation layer, requires no PMU hardware modification, and preserves full IEEE C37.118 compatibility. To assess robustness beyond a static attack pool, we further cast ROAM's defense as a \rev{competitive} Markov game against an adaptive \ac{SAC} attacker that observes and reacts to the defender's configuration. The primary contributions are as follows.
\begin{enumerate}
\item A hybrid \ac{MTD} architecture that unifies seed and ML-prediction-based ID remapping with per-sensor mode assignment at the PDC layer, \rev{removing the single seed dependency of prior deception frameworks~\cite{ddaf21, iddaf21}: a seed compromise affects only the seed-routed subset, while the ML-routed subset is recovered without the seed.}

\item A DQN-based defense agent that \rev{selects the deception amount, per-sensor remapping flags, and the seed/ML recovery split through a factored action whose size grows only linearly with the number of PMUs, under a multi-objective reward driven by the measured detection outcome, deception cost, latency, and attacker accuracy.}

\item An adversarial co-evolution evaluation protocol in which a SAC attacker learns targeted, stealthy injection strategies conditioned on the defender's most recent action, exposing topology-dependent policy vulnerabilities that static attack scenarios cannot reveal.

\item Real-time validation via sim-to-real transfer of the trained policy, without retraining, to an OPAL-RT hardware-in-the-loop testbed (IEEE 118-bus, 19 PMUs), demonstrating sub-18\,ms \rev{average} shuffle execution and \rev{an average} reshuffle blind window \rev{of 2.6\%} of the decision cycle.
\end{enumerate}

The remainder of the paper is organized as follows. Section~\ref{sec:technical} discusses the key components that support our proposed framework. Section~\ref{sec: roam_framework} presents the ROAM architecture and the \ac{MDP} formulation of the defense problem. Section~\ref{sec:evaluation} reports the experimental validation, comparative analysis, adversarial evaluation against the adaptive attacker, and the OPAL-RT-based real-time deployment. Section~\ref{sec:conclusion} concludes and outlines future work.



\section{Technical Overview}
\label{sec:technical}

ROAM is a reinforcement-learning-based hybrid MTD technique applied at the PMU data aggregation layer. We define a threat model based on the attacker's capabilities and objectives, then design deception strategies to expose and engage the attacker accordingly.

\vspace{-0.1cm}
\begin{figure}[!ht]
    \centering
    \vspace{-2pt}
    \includegraphics[width=1\columnwidth]{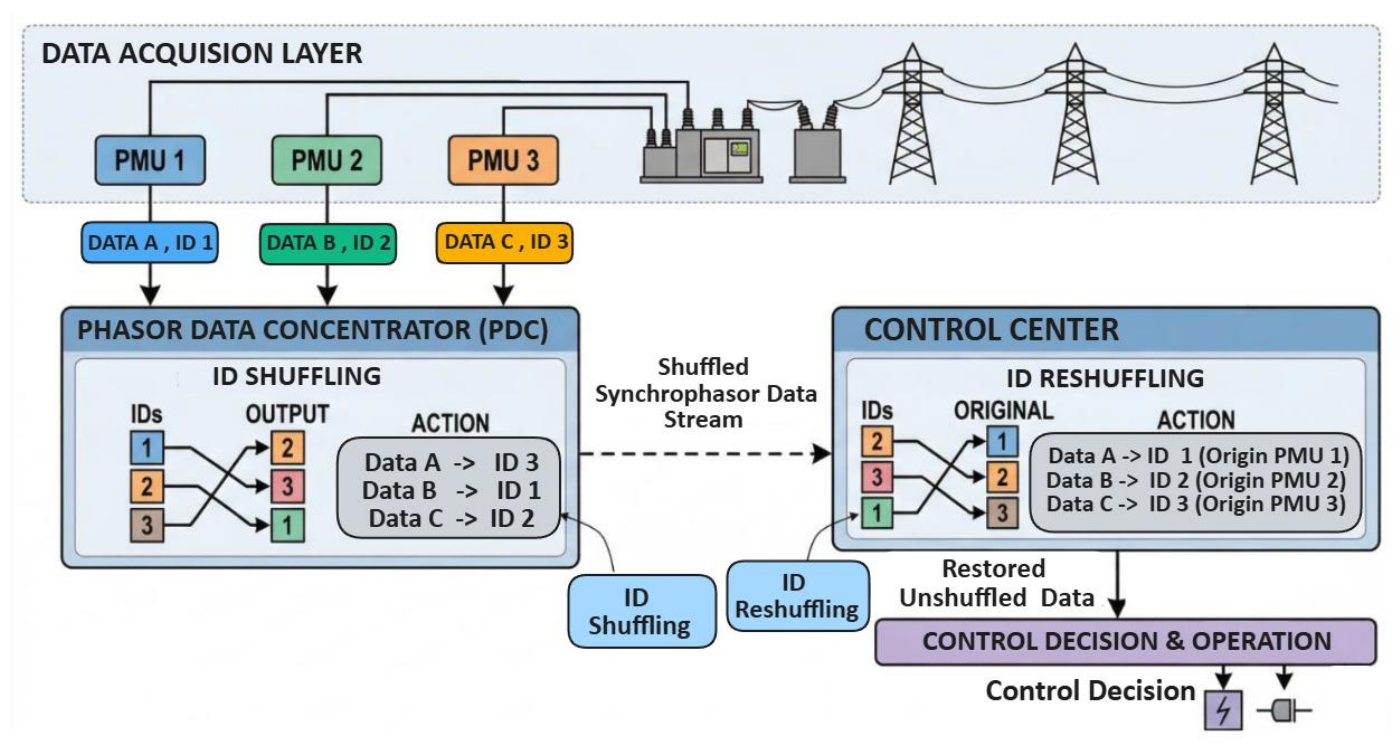}
    \vspace{-2pt}
    \caption{Proposed sensor ID shuffling and reshuffling architecture under the ROAM framework. Potential adversarial injection points are highlighted.}
    \label{fig:hybrid_mtd_update}
\vspace{-0.3cm}
\end{figure}

\subsection{Hierarchical Control Architecture}

ROAM is designed for a hierarchical control system organized across three layers, as shown in Fig.~\ref{fig:hybrid_mtd_update}. At the leaf layer, PMUs $\mathcal{P} = \{p_1, \dots, p_N\}$ generate real-time phasor measurements of bus voltages and line currents; at the aggregation layer, PDCs $\mathcal{D} = \{d_1, \dots, d_M\}$ collect and time-align data from multiple PMUs before forwarding synchronized frames to the supervisory layer; and the top-layer EMS server ingests aggregated PDC streams, executes SE and BDD, and dispatches optimal generator set points through the AGC loop. The security vulnerability is the static assignment of physical sensors to logical measurement IDs: an attacker who learns this mapping can craft FDI vectors that evade BDD while corrupting AGC frequency control. ROAM counters this by making the mapping a continuously randomized, DRL-optimized decision variable, invalidating any attacker model built on prior observations. The root of trust in ROAM's seed-based remapping pathway rests on the confidentiality and integrity of the time-varying seed $s_t$ distributed between the PDC and EMS over a hardware-backed authenticated encryption channel (AES-256-GCM over TLS~1.3) consistent with NERC CIP-007-6 requirements, with the seed generated by a hardware random-number generator (HRNG) embedded in the PDC to minimize the software attack surface.

To mitigate desynchronization risk, ROAM implements a sliding-window seed reconciliation protocol: if the EMS fails to acknowledge a seed within $\Delta_{\mathrm{sync}} = 50$\, ms, the PDC re-transmits over a secondary out-of-band channel, and the ML-based pathway is automatically activated as a fallback for all sensors in $C_t$ until synchronization is restored. Under partial key compromise, wherein an adversary obtains $k$ of $N$ seed values the effective attack accuracy degrades as:
\begin{equation}
    AA_{\mathrm{partial}} = \frac{k}{N} \cdot AA_{\mathrm{full}}
    + \left(1 - \frac{k}{N}\right) \cdot AA_{\mathrm{displaced}},
    \label{eq:partial_compromise}
\end{equation}
where $AA_{\mathrm{full}}$ is the accuracy under no defense and
$AA_{\mathrm{displaced}}$ is the residual accuracy after complete
displacement. This formulation shows that even a 50\% seed compromise yields
only partial attacker advantage, since the ML-protected sensors in
$C^{ml}_{t}$ continue to produce accurate measurements regardless of seed state. \rev{Seed compromise is not benign at every deception level: Section~\ref{sub: mtd_without_optimization} shows that a compromised seed can raise the stealthy-attack percentage above the no-deception baseline at low deception, because incorrect de-shuffling realigns falsified data with the attacker's targets. The hybrid pathway limits this exposure to the seed-routed subset $\mathcal{C}^{\mathrm{seed}}_t$, since sensors in $\mathcal{C}^{\mathrm{ml}}_t$ are recovered without the seed.}  


\subsection{State Estimation}
ROAM uses an extended Kalman filter (EKF) for recursive state
estimation, enabling continuous uncertainty propagation and real-time
anomaly detection in the adversarial setting. Let $n$ denote the state-space dimension, equal to twice the number of buses $|\mathcal{B}|$ since each bus contributes a voltage magnitude and a voltage angle: $n=2|\mathcal{B}|$. For the IEEE 14-, 57-, and 118-bus systems, this yields \rev{$n \in \{28, 114, 236\}$}, respectively. The measurement model can be described as: 
\begin{equation}
  \mathbf{z}_t = \bigl|\mathbf{H}\,\mathbf{x}_t\bigr| + \boldsymbol{\epsilon}_t,
  \quad \boldsymbol{\epsilon}_t \sim \mathcal{N}(\mathbf{0},\,\mathbf{R}), \quad \mathbf{R} = 0.01\mathbf{I}_m
  \label{eq:meas}
\end{equation}
where $\mathbf{z}_t \in \mathbb{R}^m$ is the PDC-synchronized
measurement vector, $\mathbf{x}_t \in \mathbb{R}^n$ is the bus state vector (voltage angles and magnitudes), $\mathbf{H} \in \mathbb{R}^{\rev{m\times n}}$ is the network
topology (measurement Jacobian) matrix, $\mathbf{I}_m$ is the \rev{$m\times m$} identity matrix, and $m$ is the number of PMU measurements ($m \in \{55, 218, 491\}$) for the three test systems as listed in Table~\ref{tab:dqn_ppo_comparison}. Because state changes between consecutive PDC frames are small relative to the measurement noise floor, the EKF adopts an identity process model so that the prior state equals the previous estimate\rev{ (the element-wise magnitude in~\eqref{eq:meas} makes the measurement model nonlinear, so the EKF update linearizes it around $\hat{\mathbf{x}}^-_t$)}:
\begin{equation}
  \hat{\mathbf{x}}^-_t = \hat{\mathbf{x}}_{t-1}, \qquad
  \mathbf{P}^-_t = \mathbf{P}_{t-1} + \mathbf{Q}, \qquad \mathbf{Q} =10^{-3}\mathbf{I}_n
  \label{eq:ekf_pred}
\end{equation}
where \rev{$\mathbf{I}_n \in \mathbb{R}^{n\times n}$} is the identity matrix and the \rev{scalar} $10^{-3}$ is the uniform process noise variance assigned to each state component. The filter is initialized with \rev{$\hat{\mathbf{x}}_0 = \mathbf{0}_n$ and $\mathbf{P}_0=0.1\mathbf{I}_n$}. The identity process model $\hat{x}^{-}_{t} = \hat{x}_{t-1}$ is a deliberate design choice grounded in the temporal resolution of the PMU measurement pipeline: frames arrive at 30--60 samples per second under IEEE C37.118 (16.7--33.3\, ms intervals). Empirically, the normalized state change between consecutive frames in our test datasets satisfies: $    \frac{\|\hat{x}_{t} - \hat{x}_{t-1}\|_{2}}{\|\hat{x}_{t-1}\|_{2}} < 0.8\%$
%
under normal operating conditions, and remains below $3.1\%$ during simulated $N-1$ contingency events, keeping the resulting prediction residual well within the threshold. A linearized swing-equation process model would improve state-prediction accuracy during transient events but would require real-time parameter estimation of generator inertia constants, introducing a secondary attack surface; we identify physics-informed EKF process models as a direction for future work.

\subsection{Threat Model}
\label{subsec:threat_model}
We consider an informed, persistent adversary whose objective is to corrupt AGC frequency control by injecting false data that maximizes state-estimation error while evading bad-data detection (BDD), requiring the adversary to (i) obtain the BDD mechanism's parameters, (ii) access real-time measurement data, and (iii) craft altered measurements that satisfy the BDD constraints while driving the state estimate away from ground truth. While developing the threat model, we have considered the following assumptions: 

\subsubsection*{Adversary Knowledge} The attacker has complete knowledge of the BDD constraints and the generator parameters, enabling the construction of BDD-consistent attack vectors (Assumption~I).

\subsubsection*{Adversary Capability}
The attacker can access all PMU and IED measurements, including the active power $P_b$ and frequency $\omega_b$ at the generator buses, thereby affording full system observability (Assumption~II). The attack surface can span the PMU--PDC link, the PDC--EMS link, or both.

\subsubsection*{Stealth Constraints}
The attacker may inject false data continuously or selectively across multiple time steps (Assumption~III), but system-level operational limits bound the per-step injection. The maximum allowable per-step frequency deviation is governed by:
\begin{equation}
 \small -\tau_{\omega}^b \leq \bigl(\omega^b[t{+}1] + \tilde{\omega}^b[t{+}1]\bigr) - \bigl(\omega^b[t] + \tilde{\omega}^b[t]\bigr) \leq \tau_{\omega}^b, \quad \forall b \in \mathcal{B}^{GV},
  \label{eq:omega_constraint}
\end{equation}
and the analogous active-power constraint is:
\begin{equation}
 \small  -\tau_{P}^b \leq \bigl(P^b[t{+}1] + \tilde{P}^b[t{+}1]\bigr) - \bigl(P^b[t] + \tilde{P}^b[t]\bigr) \leq \tau_{P}^b, \quad \forall b \in \mathcal{B}^{GV},
  \label{eq:power_constraint}
\end{equation}
where $\omega^b[t]$, $P^b[t]$ are the true frequency and active power at bus $b$ and time $t$ over the set $\mathcal{B}^{GV}$ of generator-governor buses, $\tilde{\omega}^b[t]$, $\tilde{P}^b[t]$ are the attack-induced deviations, and $\tau_{\omega}^b$, $\tau_{P}^b$ are the system-imposed rate limits. These constraints enforce gradual, ramp-limited injection: the adversary cannot induce instantaneous disruption but must steer the system toward an unsafe state over several time steps, opening the detection window that ROAM exploits through timely reshuffling.

\subsubsection*{Formal Attack Model}
Under these assumptions, the FDI attack vector at bus $b$ and time $t$ is:
\begin{equation}
  \tilde{\mathbf{z}}^b[t] = \mathbf{z}^b[t] + \boldsymbol{\delta}^b[t], \quad \|\boldsymbol{\delta}^b[t]\|_\infty \leq \tau^b,
  \label{eq:fdi_vector}
\end{equation}
where $\mathbf{z}^b[t]$ is the true measurement vector, $\boldsymbol{\delta}^b[t]$ is the injected perturbation, and the $\ell_\infty$ bound enforces the system-level rate limits.

\subsubsection*{Trust Boundary}
The adversary is assumed unable to compromise the pre-established secure seed channel between the PDC and EMS or to modify PMU hardware; these lie outside the attack surface and form the root of trust on which seed-based remapping relies. \rev{Seed compromise is nevertheless evaluated explicitly in Section~\ref{sub: mtd_without_optimization} as a failure case.} Equations~\eqref{eq:omega_constraint}--\eqref{eq:fdi_vector} define the baseline adversary against which ROAM is designed\rev{; the adaptive attacker of Section~\ref{sec:adversarial_model}, which observes and reacts to the defender's configuration, provides a worst-case robustness evaluation.}

\rev{
Since the attacker observes measurement values (Assumption~II), it could attempt the prediction-and-matching recovery the EMS uses on its ML pathway (Section~\ref{sec:ml_remapping}). Three properties limit this: (i) the remapped and seed-routed subsets are re-drawn every epoch and unknown to the attacker, so it must match all $m$ streams rather than the EMS's reduced ML subset; (ii) recovery must finish within one PMU frame (16.7--33.3\,ms) to inject a BDD-consistent vector into that frame, whereas even the EMS's reduced-subset matching takes 17--92\,ms (Table~\ref{tab:remapping_comparison}); and (iii) any mis-assignment redirects the injection to a non-target sensor, lowering attack accuracy and breaking BDD consistency. An attacker able to recover the full mapping within one frame is outside our threat model.}

\section{ROAM Framework}
\label{sec: roam_framework}


\begin{figure*}[!ht]
\centering
    \vspace{-2pt}
    \includegraphics[width=0.90\textwidth]{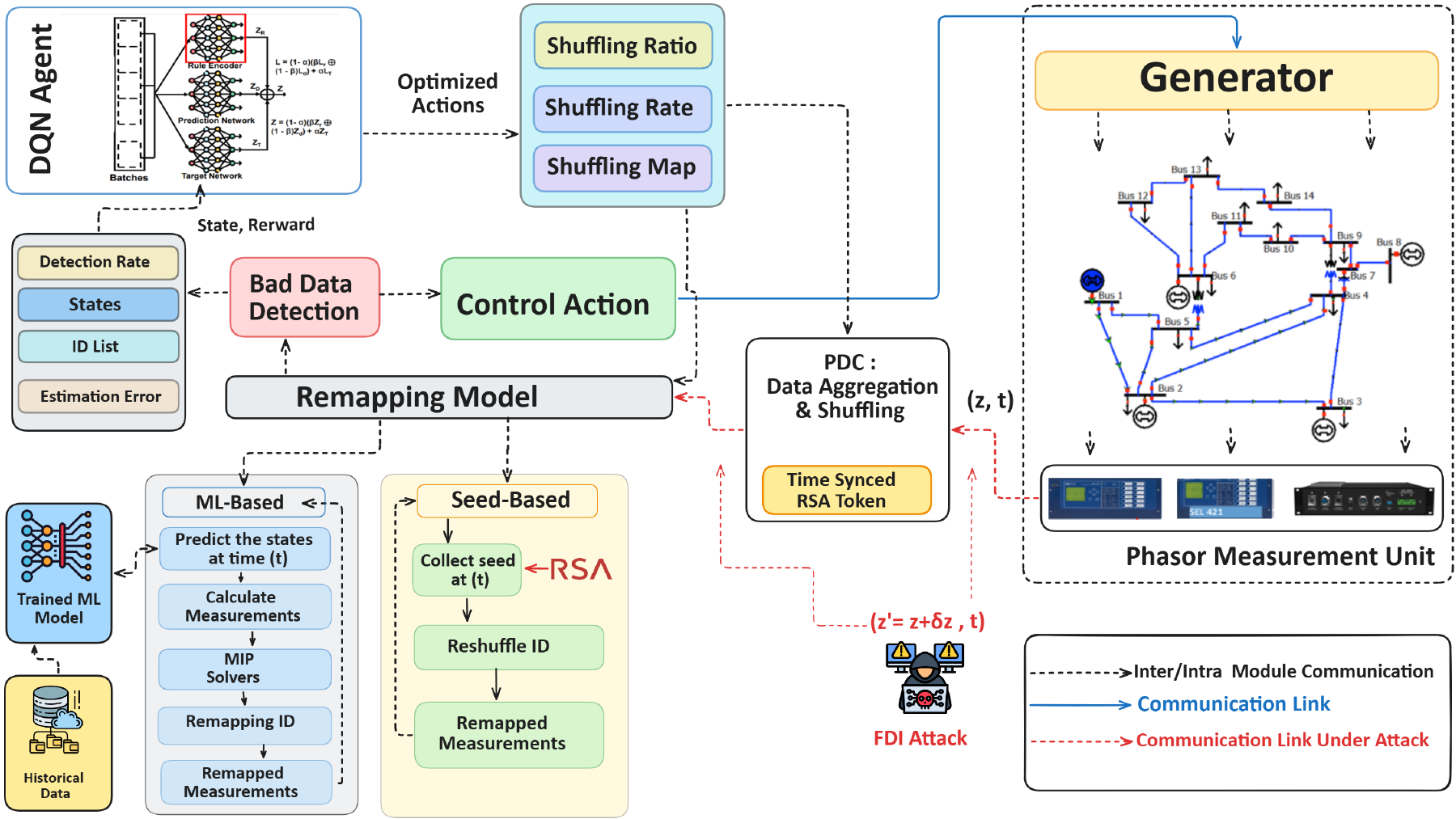}
    \vspace{-2pt}
    \caption{Proposed DQN-based hybrid MTD: the PDC shuffles PMU IDs, the EMS recovers them through seed-based or ML-based remapping before BDD, and the DQN agent adapts the shuffling ratio, rate, and map from detection and estimation feedback.}
    \label{fig:hybrid_drl_mtd}
    \vspace{-5pt}
\end{figure*}


ROAM implements MTD entirely at the aggregation layer (i.e., \rev{the PDC}), requiring no modification to PMU hardware and preserving full IEEE~C37.118 compatibility. The EMS classifies each PMU into one of two operational roles at every decision epoch:

\begin{itemize}
  \item \textbf{Fixed:} measurements forwarded to SE without alteration.
  \item \textbf{Randomized:} logical sensor IDs are shuffled before
        transmission; the EMS recovers true IDs via a shared key.
\end{itemize}
Role assignment is updated \rev{at every decision epoch by the DRL agent, whose action sets the fraction of randomized sensors and the split of their remapping pathways.} The overall data flow is illustrated in Fig.~\ref{fig:hybrid_drl_mtd}.

\subsubsection{Seed-Based ID Shuffling}

Let the set of physical PMU identifiers be
$\mathcal{ID} = \{ID_1, \ldots, ID_{\rev{N}}\}$.
A time-varying seed $s_t$ is generated via a hardware random-number
generator (analogous to \rev{hardware one-time-password tokens}) and distributed exclusively to the PDC and EMS over a pre-established secure
channel.  The PDC applies a deterministic, bijective shuffling
function $f : \mathcal{ID} \times \mathbb{Z} \to \mathcal{ID}$
to produce temporary identifiers:
\begin{equation}
  ID'_i = f(ID_i,\, s_t), \quad \forall i \in \{1,\ldots,\rev{N}\},
  \label{eq:shuffle_fn}
\end{equation}
yielding a collision-free mapping (permutation) over one shuffle
cycle.  The internal mapping table maintained by the PDC is:
\begin{equation}
  \mathcal{M}_t = \bigl\{(ID_i,\, ID'_i) :
    ID'_i = f(ID_i, s_t),\; \forall i \bigr\}.
  \label{eq:map_table}
\end{equation}
The PDC transmits the obfuscated data stream to the EMS, preserving
measurement values and timestamps while replacing identifiers:
\begin{equation}
  \mathcal{D}_t^{\mathrm{tx}}
    = \bigl\{(ID'_i,\, m_i,\, t_i) :
      (ID_i,\, m_i,\, t_i) \in \mathcal{D}_t^{\mathrm{rx}}\bigr\}.
  \label{eq:pdc_tx}
\end{equation}
Upon reception, the EMS applies the inverse function
$f^{-1}(\cdot, s_t)$ to recover authentic identities:
\begin{equation}
  ID_i = f^{-1}(ID'_i,\, s_t),
  \quad \forall ID'_i \in \mathcal{D}_t^{\mathrm{tx}}.
  \label{eq:decode}
\end{equation}
The time-varying seed prevents an adversary from inferring future
mappings from intercepted historical traffic.  This pathway yields the
seed-remapped measurement vector $\mathbf{z}_{\mathrm{seed}}(t)$.

\begin{table}[h]
\centering
\caption{ML Model Configurations in ROAM}
\label{tab:ml_models}
\begin{tabular}{p{.10\columnwidth}p{.08\columnwidth}p{.58\columnwidth}p{.05\columnwidth}}
\hline
\textbf{Model} & \textbf{Input} & \textbf{Key Hyperparameters} & \textbf{Loss} \\
\hline
XGBoost   & $\boldsymbol{\phi}_t$ & 200 trees, depth~6, lr~0.1, subsample~0.8 & MSE \\
Random Forest & $\boldsymbol{\phi}_t$ & 100 trees, depth~10 & MSE \\
LSTM      & $\mathbf{X}_{t-T:t}$ & $[64\!\to\!32]$, Dropout~0.2, Adam~$10^{-3}$, 20~ep & MSE \\
GRU       & $\mathbf{X}_{t-T:t}$ & $[64\!\to\!32]$, Dropout~0.2, Adam~$10^{-3}$, 20~ep & MSE \\
\hline
\end{tabular}
\end{table}
\subsubsection{ML-Based State Prediction and Remapping}
\label{sec:ml_remapping}
When seed synchronization is unavailable (e.g., communication link compromise) or the DRL agent \rev{selects it}, ROAM switches to a prediction-based remapping pathway: a suite of ML models, trained offline on historical state trajectories, predicts the next system state $\hat{\mathbf{x}}^-_t$ and the corresponding expected measurement vector $\tilde{\mathbf{z}}_t$, which is then used to reorder the arriving shuffled measurements $\mathbf{z}_{\mathrm{shuf}}(t)$ back to their correct sensor positions via combinatorial assignment. Two model families compete for this role, distinguished by how they consume the lookback window $\mathbf{X}_{t-T:t} \in \mathbb{R}^{T \times F}$. Tree-based models (XGBoost, Random Forest) operate on a hand-crafted feature vector $\boldsymbol{\phi}_t \in \mathbb{R}^{5F}$:
\begin{equation}
  \boldsymbol{\phi}_t =
    \bigl[\,
      \mathbf{x}_{t-1} \;\|\;
      \bar{\mathbf{x}} \;\|\;
      \boldsymbol{\sigma}_{\mathbf{x}} \;\|\;
      \mathbf{x}_{t-1} - \mathbf{x}_{t-T} \;\|\;
      \mathbf{x}_{t-1} - \mathbf{x}_{t-2}
    \,\bigr],
  \label{eq:features}
\end{equation}
where $\bar{\mathbf{x}}$ and $\boldsymbol{\sigma}_{\mathbf{x}}$ are the window mean and standard deviation, the fourth block captures the long-horizon trend, and the fifth captures the instantaneous rate of change. Sequential models (LSTM, GRU), by contrast, consume the raw scaled sequence $\mathbf{X}_{t-T:t}$ directly, without feature engineering. All four models are trained independently on an 80/20 chronological train/test split of historical state data, with configurations summarized in Table~\ref{tab:ml_models}. The LSTM and GRU networks share the same architecture, trained with mean-squared-error loss $\mathcal{L} = \frac{1}{N}\sum_t \|\mathbf{x}_t - \hat{\mathbf{x}}^-_t\|^2$ and the Adam optimizer. Each model is evaluated on the held-out test set, and the model with the lowest test MSE is selected as the active predictor:
\begin{equation}
  \mathcal{M}^* = \arg\min_{\mathcal{M}_k \in \{\text{XGB, RF, LSTM, GRU}\}}
  \frac{1}{|\mathcal{T}_{\mathrm{test}}|}
  \sum_{t \in \mathcal{T}_{\mathrm{test}}}
  \bigl\|\mathbf{x}_t - \hat{\mathbf{x}}_t^{(k)}\bigr\|^2.
  \label{eq:model_select}
\end{equation}
This selection is performed once after offline training, and the winning model is saved for deployment. Once $\tilde{\mathbf{z}}_t = |\mathbf{H}\hat{\mathbf{x}}^-_t|$ is available, recovering the true sensor ordering from $\mathbf{z}_{\mathrm{shuf}}(t)$ is cast as a binary assignment problem. Let $M_{ij} \in \{0,1\}$ indicate that shuffled measurement $j$ is assigned to predicted position $i$; the cost matrix entry is $c_{ij} = |\tilde{z}_i - z_{\mathrm{shuf},j}|$. The one-to-one matching constraints enforce
$\sum_j M_{ij} \leq 1$ and $\sum_i M_{ij} \leq 1$, while infeasible pairs exceeding a proximity threshold are suppressed before solving. This \rev{mixed-integer program} is solved with the OR-Tools CBC solver~\rev{\cite{ortools}}; the resulting permutation $\pi^*$ recovers $\hat{\mathbf{z}}_t = \mathbf{z}_{\mathrm{shuf}}[\pi^*]$, which feeds the EKF prediction step~\eqref{eq:ekf_pred} and the subsequent BDD stage.

\subsubsection{Hybrid Remapping}
\label{sec:hybrid_remapping}
The seed-based and ML-based pathways need not operate in isolation. When the DRL agent selects \rev{a seed ratio $\rho_t$ (Section~\ref{sec:mdp}), giving $(r_{\mathrm{seed}}, r_{\mathrm{ml}}) = (\rho_t/100,\, 1-\rho_t/100)$}, ROAM partitions the set of selected sensors $\mathcal{C}_t = \{i: c_t^{(i)} = 1\}$ into two disjoint subsets, one routed through each pathway:
\begin{equation}
  |\mathcal{C}_t^{\mathrm{seed}}| = \lfloor r_{\mathrm{seed}} \cdot |\mathcal{C}_t| \rfloor, \qquad
  \mathcal{C}_t^{\mathrm{ml}} = \mathcal{C}_t \setminus \mathcal{C}_t^{\mathrm{seed}},
  \label{eq:hybrid_split}
\end{equation}
where $r_{\mathrm{seed}} + r_{\mathrm{ml}} = 1$. Sensors in $\mathcal{C}_t^{\mathrm{seed}}$ are remapped via the seed-based shuffling; sensors in $\mathcal{C}_t^{\mathrm{ml}}$ are reordered using the ML prediction just described. The two recovered groups are then merged with the fixed (non-shuffled) sensors $\mathcal{F}_t$ into a single reconstructed measurement vector:
\begin{equation}
  \mathbf{z}_{\mathrm{rem}}(t) =
  \mathbf{z}_{\mathrm{seed}}(t)\big|_{\mathcal{C}_t^{\mathrm{seed}}} \cup
  \mathbf{z}_{\mathrm{ml}}(t)\big|_{\mathcal{C}_t^{\mathrm{ml}}} \cup
  \mathbf{z}(t)\big|_{\mathcal{F}_t}.
  \label{eq:hybrid_merge}
\end{equation}
If BDD confirms that $\mathbf{z}_{\mathrm{rem}}(t)$ preserves system observability, it is forwarded directly to SE. Otherwise, the EMS falls back on a weighted blend with the predicted measurements:
\begin{equation}
  \mathbf{z}_{\mathrm{fin}}(t) = \alpha\, \mathbf{z}_{\mathrm{rem}}(t) + (1-\alpha)\, \tilde{\mathbf{z}}(t), \qquad \alpha \in [0,1],
  \label{eq:hybrid_fallback}
\end{equation}
where $\alpha$ is set to $1$ when observability is satisfied and reduced toward $0$ in proportion to the fraction of flagged sensors, ensuring SE continuity even under severe disruption.

\subsubsection{DQN-Directed Adaptive Control of the Hybrid MTD}

The DQN agent closes the loop by \rev{reconfiguring the hybrid MTD} at each decision epoch. At each step $t$, the agent observes the EKF state and the BDD \rev{outcome}, then outputs \rev{$(d_t, \mathbf{c}_t, \rho_t)$}, encoding: (i) how many sensors to shuffle ($d_t$), \rev{(ii)~which sensors, through a per-sensor flag vector $\mathbf{c}_t$ that defines the remapped subset $\mathcal{C}_t$, and (iii)~how to recover them, i.e., the share $\rho_t$ routed through seed-based recovery, from mostly ML ($\rho_t$ small) to mostly seed-based ($\rho_t$ large).} The agent is rewarded \rev{by the BDD detection rate measured at that step} and penalized for deception cost, latency, and \rev{the measured} attacker accuracy \rev{(Section~\ref{sec:mdp})}. This feedback-driven, closed-loop design distinguishes ROAM from prior fixed-deception or random-segregation MTD schemes that do not learn from system feedback.

\subsection{Markov Decision Process Formulation}
\label{sec:mdp}

The ROAM defense problem is formulated as a discrete-time Markov Decision Process (MDP) $\mathcal{M} = \langle \mathcal{S}, \mathcal{A}, \mathcal{T}, \mathcal{R}, \gamma \rangle$, where $\mathcal{S}$ is the state space, $\mathcal{A}$ is the action space, $\mathcal{T}$ is the transition kernel, $\mathcal{R}$ is the reward function, and $\gamma \in (0,1)$ is the discount factor. At each decision step $t$, the agent observes a temporal window of EKF-estimated, normalized system states: $  \mathbf{s}_t = \mathbf{X}_t \in \mathbb{R}^{T \times F}, $ where $T$ is the lookback window length (e.g., $T = 10$) and $F$ is the \rev{number of monitored state features} (e.g., $F = 13$ for the IEEE 14-bus system, one per non-slack bus). Each row $\mathbf{x}_\tau \in \mathbb{R}^F$ of $\mathbf{X}_t$ is obtained from the EKF state estimate and normalized \rev{by a batch-normalization layer at the network input}:
\begin{equation}
  \hat{\mathbf{X}}_t = \gamma_{\text{bn}} \frac{\mathbf{X}_t - \mu_B}{\sqrt{\sigma_B^2 + \epsilon}} + \beta_{\text{bn}},
  \label{eq:batchnorm}
\end{equation}
where $\mu_B$, $\sigma_B^2$ are the \rev{mini-batch mean and variance during training (running averages at inference)}, $\epsilon$ is a numerical stability constant, and $\gamma_{\text{bn}},\, \beta_{\text{bn}}$ are learnable affine parameters. This normalization keeps state observations well-conditioned across varying operating conditions and attack magnitudes. The ROAM agent's action $a_t \in \mathcal{A}$ is a \rev{three-component tuple}:
\begin{equation}
  {\revstart a_t = \bigl(d_t,\; \mathbf{c}_t,\; \rho_t\bigr), \;\; d_t \in \{25, \ldots, 80\},\; \rho_t \in \{10, \ldots, 90\},}
  \label{eq:action}
\end{equation}
where \rev{$d_t$ is the deception amount, i.e., the percentage of PMUs selected for ID shuffling at step $t$, and $\rho_t$ is the seed ratio, i.e., the percentage of the selected PMUs recovered through seed-based remapping. The agent flags the sensors to shuffle through the binary sensor-selection vector $\mathbf{c}_t \in \{0,1\}^N$, constrained by $\sum_{i=1}^{N} c_t^{(i)} = \lfloor N d_t/100 \rfloor$; for example, with $d_t = 50$ and $N = 100$, it flags exactly which 50 PMUs are remapped. The seed ratio sets $r_{\mathrm{seed}} = \rho_t/100$ in~\eqref{eq:hybrid_split}, so every selected sensor is assigned its own recovery mode: seed-based for $\mathcal{C}^{\mathrm{seed}}_t$ and ML-based for $\mathcal{C}^{\mathrm{ml}}_t$.}

When the EMS receives measurements with shuffled IDs, it routes each sensor through its assigned remapping process, reconstructs the measurement vector \rev{via~\eqref{eq:hybrid_merge}}, and forwards the combined output to the SE-BDD pipeline. The environment transitions according to:
\begin{equation}
  P(\mathbf{s}_{t+1} \mid \mathbf{s}_t, a_t) = P\!\left(\mathbf{X}_{t+1} \mid \mathbf{X}_t,\, {\revstart d_t,\, \mathbf{c}_t,\, \rho_t}\right),
  \label{eq:transition}
\end{equation}
capturing stochasticity arising from adversarial FDI injections, network measurement noise, and the effectiveness of the current defense configuration. The composite reward function guides the ROAM agent to simultaneously maximize attack \rev{detection}, minimize operational overhead, and suppress attacker accuracy. It is defined as:
\begin{equation}
  {\revstart r_t = \lambda_3 D_t - \lambda_1 C^{\text{dec}}_t - \lambda_2 L_t - \lambda_4 A_t,}
  \label{eq:reward}
\end{equation}
\rev{where $D_t$ is the BDD detection rate (CSER, Section~\ref{sub: mtd_without_optimization}) measured on the attacked frames of step $t$, and $A_t$ is the attack accuracy (AA) measured at step $t$, i.e., the fraction of injections that still land on their intended sensors after remapping. Both terms therefore depend on the state, the injected attack, and the realized remapping, not on the action alone. The deception cost $C^{\text{dec}}_t = d_t/100$ penalizes the operational overhead of shuffling and the latency overhead: }
\begin{equation}
  {\revstart L_t = \frac{d_t}{100}\Bigl(\frac{\rho_t}{100}\,\kappa_1 + \Bigl(1-\frac{\rho_t}{100}\Bigr)\kappa_2\Bigr)}
  \label{eq:latency}
\end{equation}
\rev{captures the processing delay of the remapping mix,} with $\kappa_2 > \kappa_1$ reflecting the higher computational cost of ML-based remapping relative to a simple shuffle. \rev{We selected the weights $\{\lambda_1, \lambda_2, \lambda_3, \lambda_4, \kappa_1, \kappa_2\}$ via a grid search over the IEEE 14-bus system and held them fixed across all topologies to ensure a fair cross-system comparison.} The two most influential parameters are $\lambda_3$ (detection scaling) and $\lambda_4$ (attacker accuracy penalty), which jointly control the fundamental security-vs-overhead trade-off.

To assess sensitivity, we varied $\lambda_3, \lambda_4 \in \{0.5, 1.0, 2.0, 4.0\}$ independently across 16 configurations on the IEEE 14-bus system over 50 held-out episodes. The mean PSA ranged from $1.8\%$ (at $\lambda_3=\lambda_4=4.0$) to $5.2\%$ (at $\lambda_3=\lambda_4=0.5$), confirming the policy is robust to moderate perturbations but degrades predictably when detection incentives are under-weighted. \rev{Because changing $\lambda_1$ or $\lambda_2$ shifts the optimal trade-off, operators who recalibrate them for system-specific latency budgets should fine-tune the trained policy rather than reuse it unchanged.}

\noindent\textbf{Action Space \rev{Size}.}
\rev{The action is factored: the deception level and seed ratio form $12 \times 5 = 60$ discrete choices, and the sensor subset is expressed through $N$ per-sensor flags rather than by enumerating subsets. Enumerating the subsets jointly would require $\sum_{d}\binom{N}{\lfloor N d/100\rfloor}$ choices (about $5.2\times 10^{5}$ for $N=19$), whereas the flag representation grows only linearly in $N$; the per-epoch remapping and ML-matching cost also grows with $N$ (Section~\ref{sub: mtd_without_optimization}).}

\begin{table}[htbp]
\centering
\caption{Deep Q-Network (DQN) Hyperparameter Configuration}
\label{tab:dqn_hyperparameters}
\begin{tabular}{p{0.30\columnwidth}p{0.30\columnwidth}p{0.25\columnwidth}}
\hline
\textbf{Category} & \textbf{Hyperparameter} & \textbf{Value} \\
\hline
\multirow{2}{*}{Optimization} 
    & Learning rate ($\alpha$) & $5 \times 10^{-4}$ \\
    & Discount factor ($\gamma$) & 0.99 \\
\hline
\multirow{3}{*}{Experience Replay} 
    & Buffer size & 100,000 \\
    & Batch size & 128 \\
    & Learning starts & 1,000 \\
\hline
\multirow{2}{*}{Target Network} 
    & Soft update coefficient ($\tau$) & 0.001 \\
    & Hard update interval & 1,000 steps \\
\hline
\multirow{2}{*}{Training Schedule} 
    & Train frequency & 4 steps \\
    & Gradient steps & 1 \\
\hline
\multirow{3}{*}{Exploration} 
    & Initial $\epsilon$ & 1.0 \\
    & Final $\epsilon$ & 0.05 \\
    & Exploration fraction & 0.2 \\
\hline
\multirow{2}{*}{Network Architecture} 
    & Hidden layers & $[256  \to 512 \to 512 \to 256]$ \\
    & Activation function & ReLU \\
\hline
Training Duration & Total timesteps & 40,000 \\
\hline
\end{tabular}
\end{table}


\subsection{Primary DRL Model}

ROAM employs a deep Q-network (DQN) as its primary defense agent, motivated by the discrete, combinatorial nature of the action space in~\eqref{eq:action}: the deception amount $d_t$ \rev{and the seed ratio $\rho_t$ are drawn from finite sets of integer percentages, and the sensor flags are binary}. DQN's explicit Q-value enumeration over this discretized space provides a direct, interpretable mechanism for selecting optimal shuffle configurations at each step. The DQN agent approximates the optimal action-value function via a deep neural network $Q_\theta(\mathbf{s}_t, a_t)$, updated via the Bellman target:
\begin{equation}
  y_t^{\text{DQN}} = r_t + \gamma \max_{a'} Q_{\theta^-}\!\left(\mathbf{s}_{t+1}, a'\right),
  \label{eq:bellman}
\end{equation}
where $\theta^-$ are the parameters of a periodically synchronized target network, decoupling the bootstrapping target from the online network to reduce oscillations. The online network is updated by minimizing the temporal-difference loss:
\begin{equation}
  \mathcal{L}(\theta) = \mathbb{E}_{(\mathbf{s}_t,a_t,r_t,\mathbf{s}_{t+1})\sim\mathcal{B}} \left[ \left( y_t^{\text{DQN}} - Q_\theta(\mathbf{s}_t, a_t) \right)^2 \right],
  \label{eq:td_loss}
\end{equation}
where $\mathcal{B}$ is an experience replay buffer of capacity $|\mathcal{B}| = 10^5$ transitions. The optimal defense policy is then recovered as:
\begin{equation}
  \pi^*_{\text{DQN}}(\mathbf{s}_t) = \arg\max_{a_t \in \mathcal{A}}\, Q_\theta(\mathbf{s}_t, a_t).
  \label{eq:dqn_policy}
\end{equation}

The Q-network is a four-layer MLP with ReLU activations and a symmetric expansion–contraction structure, extracting high-dimensional feature representations before compressing them into action-value estimates, well-suited to the temporal dependencies in power system dynamics. The key hyperparameters of the DQN model are summarized in Table~\ref{tab:dqn_hyperparameters}.


\begin{table}[htbp]
\centering
\caption{Proximal Policy Optimization (PPO) Hyperparameter Configuration}
\label{tab:ppo_hyperparameters}
\begin{tabular}{p{0.30\columnwidth}p{0.40\columnwidth}p{0.15\columnwidth}}
\hline
\textbf{Category} & \textbf{Hyperparameter} & \textbf{Value} \\
\hline
\multirow{2}{*}{Optimization} 
    & Learning rate ($\alpha$) & $3 \times 10^{-4}$ \\
    & Discount factor ($\gamma$) & 0.99 \\
\hline
\multirow{3}{*}{Rollout Collection} 
    & Steps per rollout ($n_{\text{steps}}$) & 2,048 \\
    & Batch size & 64 \\
    & Number of epochs & 10 \\
\hline
\multirow{2}{*}{Advantage Estimation} 
    & GAE lambda ($\lambda$) & 0.95 \\
    & Normalize advantage & True \\
\hline
Policy Clipping & Clip range ($\epsilon_{\text{clip}}$) & 0.2 \\
\hline
\multirow{3}{*}{Loss Function Weights} 
    & Entropy coefficient ($c_{\text{ent}}$) & 0.0 \\
    & Value function coefficient ($c_{\text{vf}}$) & 0.5 \\
    & Max gradient norm & 0.5 \\
\hline
\multirow{2}{*}{Network Architecture} 
    & Policy network & MlpPolicy \\
    & Activation function & Tanh \\
\hline
\end{tabular}
\end{table}

\subsection{Comparative Baseline}

To validate the design choices of ROAM-DQN, we develop a parallel PPO-based defender that solves the same MDP \rev{(Section~\ref{sec:mdp})} via an on-policy policy-gradient approach, thereby providing a controlled comparison between off-policy value-based and on-policy methods for the MTD problem. PPO~\rev{\cite{GAE}} optimizes a clipped surrogate objective to prevent destructively large policy updates:
\begin{equation}
  \small \mathcal{L}^{\text{CLIP}}(\theta) = \mathbb{E}_t \left[
    \min\!\left(
      \rho_t(\theta)\, \hat{A}_t,\;
      \mathrm{clip}\!\left(\rho_t(\theta), 1-\varepsilon_{\text{clip}}, 1+\varepsilon_{\text{clip}}\right) \hat{A}_t
    \right)
  \right],
  \label{eq:ppo_clip}
\end{equation}
where $\rho_t(\theta) = \pi_\theta(a_t \mid \mathbf{s}_t) / \pi_{\theta_{\text{old}}}(a_t \mid \mathbf{s}_t)$ is the probability ratio between the updated and old policies, $\hat{A}_t$ is the Generalized Advantage Estimate (GAE), and $\varepsilon_{\text{clip}} = 0.2$ implements the trust-region constraint. The full PPO objective augments~\eqref{eq:ppo_clip} with a value function loss and an entropy bonus:
\begin{equation}
  \mathcal{L}^{\text{PPO}}(\theta) = \mathcal{L}^{\text{CLIP}} - c_{\text{vf}} \mathcal{L}^{\text{VF}} + c_{\text{ent}} \mathcal{H}[\pi_\theta],
  \label{eq:ppo_full}
\end{equation}
where $c_{\text{vf}} = 0.5$ and the entropy coefficient $c_{\text{ent}}$ encourages policy diversity. Advantage estimates use GAE~\rev{\cite{GAE_schulman}} with $\lambda_{\text{GAE}} = 0.95$ for low-bias, moderate-variance estimation of cumulative attack consequences. PPO \rev{shares DQN's observation, action space, and reward to ensure a fair comparison; its network and key parameters are listed in Table~\ref{tab:ppo_hyperparameters}.}

\subsection{Adversarial Multi-Agent Framework}
\label{sec:adversarial_model}

To stress-test ROAM under worst-case conditions beyond the static scenario pool, we formulate an adversarial evaluation framework as a \rev{competitive} Markov game between the trained defender (DQN or PPO) and an adaptive attacker. Formally, the game is defined as $\mathcal{G} = \langle \mathcal{S}, \mathcal{A}_{\text{def}}, \mathcal{A}_{\text{atk}}, \mathcal{T}, \mathcal{R}_{\text{def}}, \mathcal{R}_{\text{atk}}, \gamma \rangle$, \rev{where the defender keeps its reward~\eqref{eq:reward} and the attacker's reward~\eqref{eq:atk_reward} directly opposes its detection and state-accuracy terms.}

The adaptive attacker is modeled as a continuous-action RL agent trained with Soft Actor-Critic (SAC)~\cite{haarnoja2018soft}. SAC is selected for its entropy-regularized policy:
\begin{equation}
  \pi^*_{\text{atk}} = \arg\max_{\pi} \mathbb{E}_{\pi}\!\left[\sum_t \gamma^t \bigl(R_{\text{atk},t} + \alpha_{\text{ent}} \mathcal{H}[\pi(\cdot \mid \mathbf{s}_t)]\bigr)\right],
  \label{eq:sac_objective}
\end{equation}
where $\alpha_{\text{ent}}$ is the temperature parameter controlling exploration–exploitation balance. The entropy term discourages premature convergence to a single attack pattern, producing a diverse, stealthy attacker representative of a sophisticated adversary. The attacker observes the same power system state sequence as the defender, augmented with the defender's most recent action \rev{$(d_t, \rho_t)$}:
\begin{equation}
  \mathbf{o}_{\text{atk}} = \bigl[\,\mathrm{vec}(\mathbf{s}_{t-L+1:t}) \;\|\; d_t/100 \;\|\; \rev{\rho_t}/100\,\bigr] \in \mathbb{R}^{L \cdot F + 2},
  \label{eq:atk_obs}
\end{equation}
where $L$ is the lookback window and $F$ is the per-step feature dimension. This allows the attacker to condition its strategy on the defender's most recent configuration. The attacker's continuous action is a $(2|\mathcal{Z}|+1)$-dimensional vector:


%
\begin{equation}
  \mathbf{a}_{\text{atk}} = \bigl[\, \boldsymbol{\ell} \in [-1,1]^{|\mathcal{Z}|} \;\|\; \mathbf{m} \in [-1,1]^{|\mathcal{Z}|} \;\|\; \alpha \in [-1,1] \,\bigr],
  \label{eq:atk_action}
\end{equation}
where $\boldsymbol{\ell}$ are sensor-selection logits, $\mathbf{m}$ are per-sensor injection magnitude parameters, and $\alpha$ is a global intensity scalar. This is converted to a physical FDI attack matrix $\mathbf{A} \in \mathbb{R}^{|\mathcal{Z}| \times 3}$ (sensor ID, attack flag, injection value) via:
\begin{align}
  p_i &= \sigma(3\ell_i), \label{eq:atk_prob} \\
  k &= \mathrm{clip}\!\left(\textstyle\sum_i \mathbf{1}[p_i > 0.3],\; 3,\; 30\right), \label{eq:atk_count} \\
  a_i &= m_i \cdot 50 \cdot \tilde{\alpha}, \quad \tilde{\alpha} = \mathrm{clip}\!\left(\tfrac{\alpha+1}{2}, 0.1, 1.0\right), \label{eq:atk_magnitude}
\end{align}
selecting the top-$k$ sensors with $p_i > 0.3$ and injecting values in $[-50, +50]$ per sensor, scaled by the learned global intensity $\tilde{\alpha}$.

\paragraph{Attacker Reward.}
The attacker's multi-objective reward is:
\begin{equation}
  \small R_{\text{atk}} = 100 \cdot \Bigl(
    \underbrace{2.0 \cdot \tfrac{\mathrm{EDR}}{100}}_{\text{state damage}}
    - \underbrace{1.5 \cdot \tfrac{\mathrm{CSER}}{100}}_{\text{detection penalty}}
    + \underbrace{1.0 \cdot \tfrac{\mathrm{AccAtk}}{100}}_{\text{targeting accuracy}}
    - \underbrace{0.3 \cdot \tfrac{n_{\text{atk}}}{30}}_{\text{sparsity cost}}
  \Bigr)
  \label{eq:atk_reward}
\end{equation}
where EDR (Estimation Deviation Rate) quantifies damage to the SE output, CSER (\rev{Compromised Sensor Elimination Rate}) penalizes attacks that are detected, AccAtk is the fraction of injections that land on the intended targets, and $n_{\text{atk}}$ is the number of attacked sensors. The sparsity cost discourages indiscriminate multi-sensor attacks, forcing the attacker to develop targeted, stealthy strategies. \rev{Its detection and damage terms mirror the defender's $D_t$ and SE objectives with opposite sign, so the two agents are in direct competition.}

The adversarial training follows a four-phase alternating protocol: \rev{the} pre-trained DQN and PPO defenders (from standard single-agent training) are loaded with fixed weights $\theta^*_{\text{DQN}}$, $\theta^*_{\text{PPO}}$. The SAC attacker is trained for 10,000 timesteps against the frozen DQN defender to establish a meaningful baseline attack policy. Both DQN and PPO defenders are fine-tuned against the frozen pre-trained attacker, adapting their policies to the adversary's learned strategies. Finally, the SAC attacker is retrained against the updated frozen DQN defender, exploiting any new weaknesses introduced by fine-tuning.
This alternating co-evolution ensures that both the attacker and defender continuously improve relative to each other, analogous to a generative adversarial training paradigm for security policies.

We evaluate both DQN and PPO defenders over 50 episodes against the same fully trained SAC attacker, using deterministic policies ($\text{deterministic}=\mathrm{True}$) for all agents. Metrics logged per episode include PSA, PEA, CSER, EDR, attack accuracy, and cumulative reward \rev{(defined in Section~\ref{sub: mtd_without_optimization})}. This ensures a controlled, symmetric comparison of ROAM-DQN versus ROAM-PPO under identical adversarial pressure.

\section{Numerical Case Studies}
\label{sec:evaluation}

We conducted several studies to evaluate our DQN-based hybrid MTD and validate its effectiveness in a real-time hardware-in-the-loop setup. All DRL models are trained on 1,000 and evaluated on another 500 BDD-compliant FDI attack scenarios using Python 3.11 with Stable Baselines3. Real-time ID shuffling is validated \rev{with virtual and physical PMUs} via a customized PDC module over IEEE C37.118. The experimental study is organized around the following research questions (RQs):

\begin{enumerate}[leftmargin=*, label=\textbf{RQ\arabic*.}]
    \item  Under dynamic FDI attack conditions, does the hybrid remapping strategy outperform individual seed-based or ML-based approaches in attack displacement, detection rate, and state estimation accuracy? (Section~\ref{sub: mtd_without_optimization})
    
    \item When equipped with per-sensor mode assignment, does the DQN agent learn a more adaptive defense policy than a fixed single-mode strategy? (Section~\ref{sb: hybrid_mode})
    
    \item How do DQN and PPO differ in training efficiency, converged policy behavior, and defense effectiveness across IEEE 14-, 57-, and 118-bus systems? (Section~\ref{sec:comparative_analysis})
    
    \item Under held-out deterministic evaluation \rev{and against an adaptive SAC attacker}, do DQN and PPO maintain their relative defense performance, and does either algorithm consistently dominate across all grid topologies? (Sections~\ref{sec:post_training} and~\ref{sec:adversarial})
    
    \item Is \rev{ROAM} feasible in a real-time system? (Section~\ref{sec: real_time})
    
\end{enumerate}

\begin{table*}[!h]
\centering
\caption{Comparative Performance of Remapping Strategies Under Cross-Level Shuffling (IEEE 14-Bus System)}
\label{tab:remapping_comparison}
\renewcommand{\arraystretch}{1.20}
\setlength{\tabcolsep}{3.5pt}
\footnotesize
\begin{tabular}{|l|c|c|c|c|c|c|c|c|}
\hline
\textbf{Strategy} & \textbf{Dec. (\%)} & \textbf{PSA (\%)} & \textbf{PEA (\%)} & \textbf{CSER (\%)} & \textbf{EDR (\%)} & \textbf{AA (\%)} & \textbf{$T_{\text{shuf}}$ (ms)} & \textbf{$T_{\text{rec}}$ (ms)} \\
\hline\hline
\multirow{6}{*}{\rotatebox{90}{\makecell{\textbf{DDAF~\cite{ddaf21}}\\\textbf{(Seed, Same)}}}}
 & 0   & 30.63 & 69.37 & 69.37 & 31.05 & 100.00 & 0.016 & 1.52 \\
 & 20  & 12.74 & 87.26 & 79.78 & 57.61 &  62.86 & 0.017 & 1.49 \\
 & 40  &  5.37 & 94.63 & 84.47 & 54.68 &  34.77 & 0.017 & 1.50 \\
 & 60  &  3.00 & 97.00 & 84.19 & 47.84 &  19.37 & 0.017 & 1.54 \\
 & 80  &  2.27 & 97.73 & 78.22 & 29.84 &  10.16 & 0.017 & 1.50 \\
 & 100 &  0.94 & 99.06 & 84.78 & 21.09 &   6.01 & 0.017 & 1.53 \\
\hline
\multirow{6}{*}{\rotatebox{90}{\makecell{\textbf{DDAF~\cite{ddaf21}}\\\textbf{(Seed, Comp.)}}}}
 & 0   & 30.63 & 69.37 & 69.37 & 31.05 & 100.00 & 0.016 & 1.52 \\
 & 20  & 46.19 & 53.81 & 25.94 &  62.09 &  62.86  & 0.024 & 1.725 \\
 & 40  & 28.97 & 71.03 & 16.60 &  64.25 &  34.77  & 0.027 & 2.877\\
 & 60  & 17.94 & 82.06 &  7.18 &  77.49 &  19.37  & 0.026 & 3.824 \\
 & 80  &  8.68 & 91.32 & 14.97 &  97.84 &  10.16  & 0.024 & 5.175 \\
 & 100 &  5.51 & 94.49 &  8.45 & \textbf{100.00} & 6.01  & 0.024 & 6.092 \\
\hline
\multirow{6}{*}{\rotatebox{90}{\makecell{\textbf{iDDAF~\cite{iddaf21}}\\\textbf{(ML, MIP)}}}}
 & 0   & 30.63 & 69.37 & 69.37 & 31.05 & 100.00 & 0.014 & 91.91 \\
 & 20  & 15.66 & 84.34 & 75.50 & 34.11 &  62.86 & 0.013 & 92.29 \\
 & 40  &  8.94 & 91.06 & 74.69 & 31.71 &  34.77 & 0.014 & 88.82 \\
 & 60  &  4.60 & 95.40 & 76.08 & 30.37 &  19.37 & 0.014 & 88.87 \\
 & 80  &  2.38 & 97.62 & 75.87 & 23.52 &  10.16 & 0.016 & 89.65 \\
 & 100 &  1.96 & 98.04 & 66.94 & 13.07 &   6.01 & 0.015 & 90.86 \\
\hline
\multirow{6}{*}{\rotatebox{90}{\textbf{Hybrid (60/40)}}}
 & 0   & 30.63 & 69.37 & 69.37 & 31.05 & 100.00 & 0.014 & 16.99 \\
 & 20  & 11.62 & 88.38 & 81.68 & 56.08 &  62.86 & 0.013 & 17.04 \\
 & 40  &  8.31 & 91.69 & 75.87 & 37.18 &  34.77 & 0.014 & 16.06 \\
 & 60  &  3.68 & 96.32 & 81.35 & 38.01 &  19.37 & 0.014 & 16.33 \\
 & 80  &  2.68 & 97.32 & 72.86 & 22.17 &  10.16 & 0.013 & 16.68 \\
 & 100 &  1.27 & 98.73 & 78.64 & 15.67 &   6.01 & 0.014 & 17.99 \\
\hline
\end{tabular}
\end{table*}

\subsection{Evaluation of Hybrid MTD in Dynamic Attack Conditions}
\label{sub: mtd_without_optimization}

\noindent\textbf{Definition of Evaluation Metrics.}
Let $\mathcal{A}_{\text{int}}$ denote the set of intended attack targets, $\mathcal{A}_{\text{act}}$ the set of sensors actually hit after shuffling, and $\mathcal{D}_{\text{BDD}}$ the set of sensors flagged by BDD.

Attack Accuracy (AA), the fraction of injected attacks reaching their intended sensors despite ID shuffling, is: $$\text{AA} = \frac{|\mathcal{A}_{\text{int}} \cap \mathcal{A}_{\text{act}}|}{|\mathcal{A}_{\text{int}}|} \times 100\%.$$Compromised Sensor Elimination Rate (CSER), the fraction of actually-hit sensors that BDD detects, is: $$\text{CSER} = \frac{|\mathcal{D}_{\text{BDD}} \cap \mathcal{A}_{\text{act}}|}{|\mathcal{A}_{\text{act}}|} \times 100\%.$$ Percentage of Stealthy Attacks (PSA) combines attack accuracy with the detection miss rate:
$$\text{PSA} = \text{AA} \times \left(1-\frac{\text{CSER}}{100}\right)
.$$ 

Percentage of Exposed Attacks (PEA) is the complement, $\text{PEA} = 100\% - \text{PSA}$. Estimation Deviation Rate (EDR) measures the relative state estimation
error at the intended target positions, with $\mathbf{z}_{\text{rec}}$ and
$\mathbf{z}_{\text{org}}$ the recovered and original measurement vectors:

$$
  \text{EDR} = \frac{\|\mathbf{z}_{\text{rec}}[\mathcal{A}_{\text{int}}] - \mathbf{z}_{\text{org}}[\mathcal{A}_{\text{int}}]\|}
               {\|\mathbf{z}_{\text{org}}[\mathcal{A}_{\text{int}}]\|} \times 100\%
$$

\begin{figure*}[!ht]
    \centering
        \subfigure[]{
            \label{action_hybrid}
            \includegraphics[width=0.60\columnwidth]{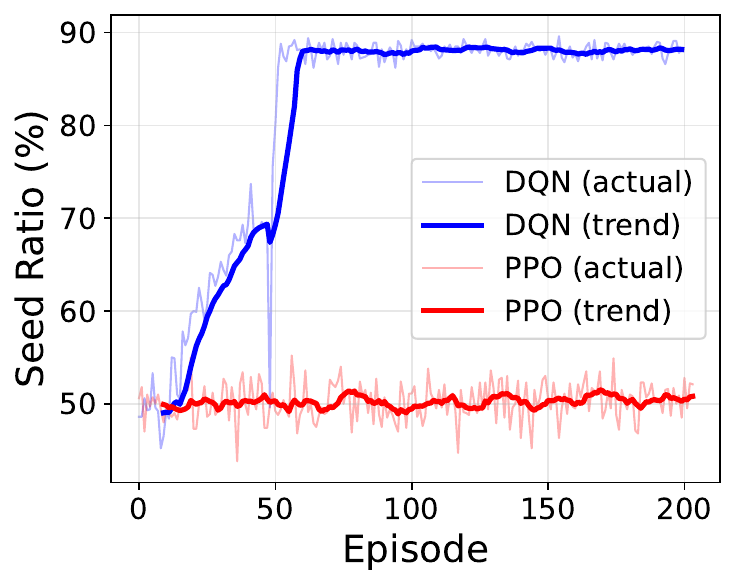}
        }
        \subfigure[]{
            \label{deception_hybrid}
            \includegraphics[width=0.60\columnwidth]{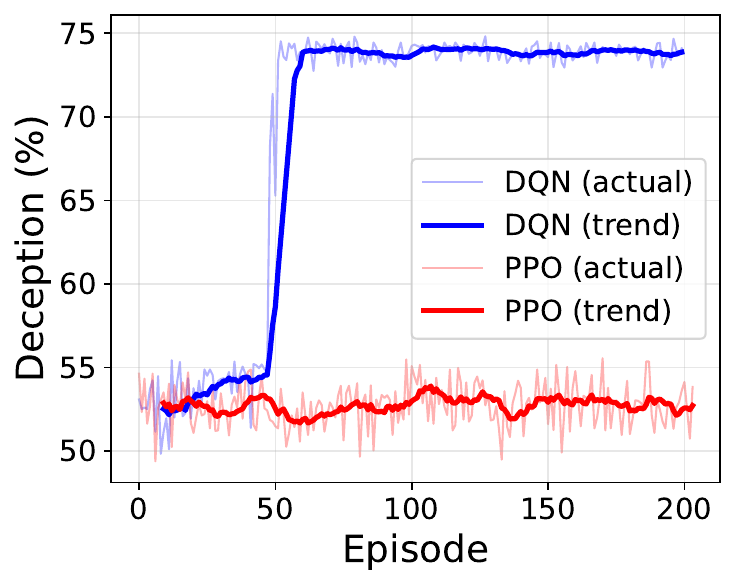}
        }
        \subfigure[]{
            \label{rewards_hybrid}
            \includegraphics[width=0.60\columnwidth]{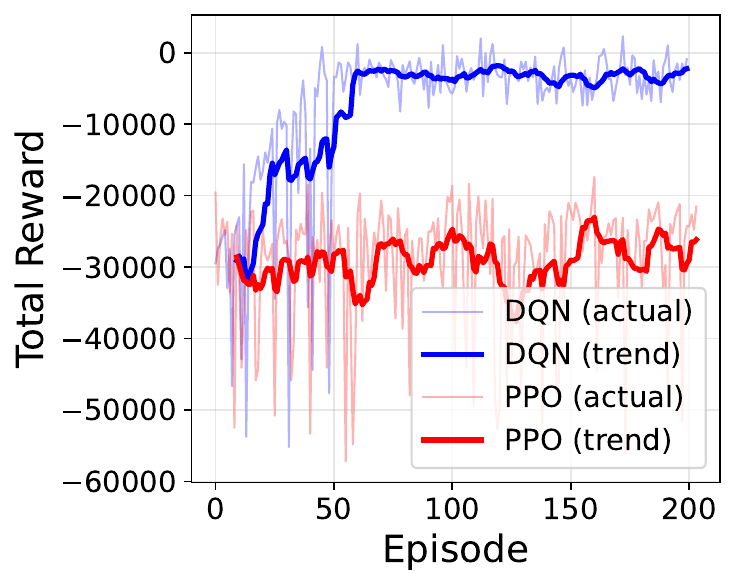}
        }\\
        \subfigure[]{
            \label{deviations_hybrid}
            \includegraphics[width=0.60\columnwidth]{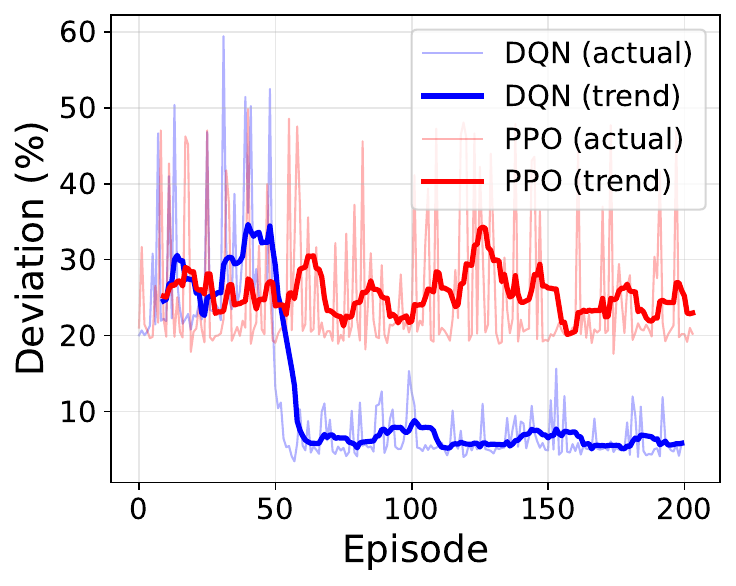}
            } 
        \subfigure[]{
            \label{detection_hybrid}
            \includegraphics[width=0.60\columnwidth]{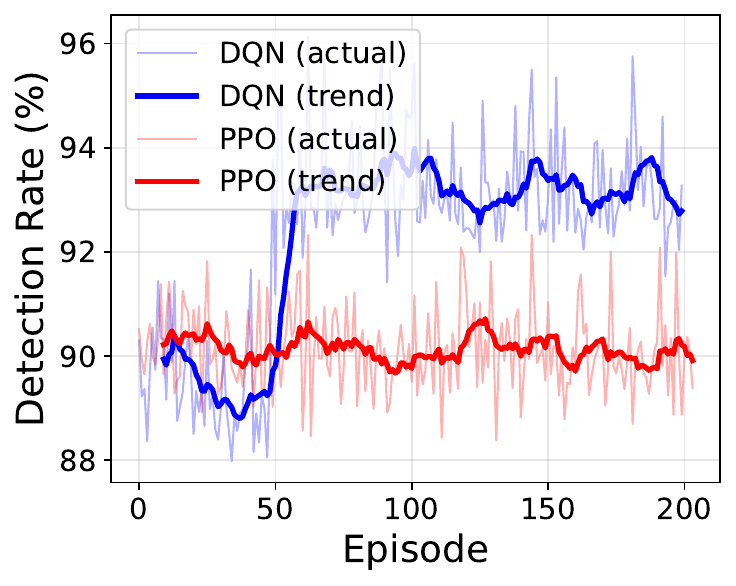}
        }
        \subfigure[]{
            \label{attack_accuracy_hybrid}
            \includegraphics[width=0.60\columnwidth]{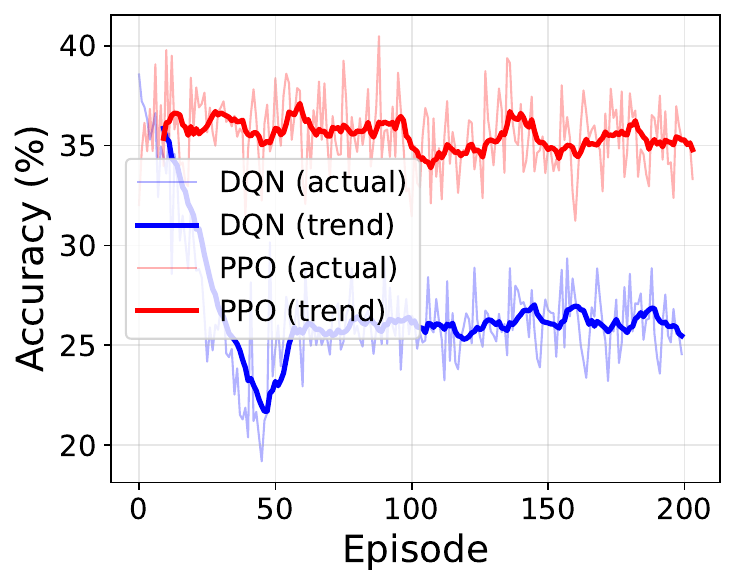}
        }
        \vspace{-6pt}
        \caption{\rev{Training behavior of ROAM-DQN and the PPO baseline in hybrid MTD mode (IEEE 118-bus system).}
(a)~\rev{Seed ratio $\rho_t$ (\%), i.e., share of remapped sensors recovered via the seed; DQN converges near 88\%.}
(b)~Deception coverage $d_t$ (\%)\rev{; DQN converges} near 75\%.
(c)~Cumulative reward convergence.
(d)~\rev{State deviation (EDR, \%); DQN settles} below 10\%.
(e)~Detection rate \rev{(CSER, \%); DQN settles} above 90\%.
(f)~Attack accuracy (\%)\rev{; DQN falls} below 30\%.}
        \label{plot:DQN performance}
\end{figure*}

Table~\ref{tab:remapping_comparison} evaluates four remapping strategies on the IEEE 14-bus system under cross-level shuffling, where sensors may swap across hierarchical boundaries to maximize displacement. The seed-based and ML-based strategies correspond to the existing DDAF~\cite{ddaf21} and iDDAF~\cite{iddaf21} frameworks, respectively, and serve as the baselines for ROAM's hybrid remapping. Attack accuracy drops monotonically from 100\% to 6.01\% at full deception across all strategies, since displacement is independent of the recovery method; the strategies differ instead in how well they reconstruct the shuffled measurements and in how much they degrade under seed compromise. Seed-based recovery achieves the strongest detection when the key remains secure, with PSA falling to 0.94\% and sub-2\,ms recovery, well-suited to real-time SCADA environments. However, it presents a single point of failure: seed compromise causes CSER to collapse \rev{(8.45\% at full deception)} and EDR to reach 100\% at full deception, and incorrect de-shuffling at \rev{low deception} can even realign attacked measurements with their intended targets, pushing PSA above the no-deception baseline.

A counterintuitive result warrants discussion: under seed compromise, PSA rises \rev{to 46.19\% at 20\% deception, above the 30.63\% no-deception baseline, and remains near it (28.97\%) at 40\%.} This occurs because incorrect de-shuffling, driven by the attacker's knowledge of the compromised seed, reassigns attacked measurements to sensor positions that coincide with the attacker's intended injection targets. The PDC's own remapping inadvertently realigns falsified data with the intended bus states, amplifying the attack's effectiveness. This underscores that \rev{a compromised seed is most dangerous at low deception levels}: full de-shuffle failure (100\% deception, seed-compromised) still achieves only 5.51\% PSA, since \rev{the resulting displacement leaves only 6.01\% of injections on their targets}. The hybrid strategy \rev{confines} this failure mode \rev{to the seed-routed subset}, since the ML-protected subset \rev{$\mathcal{C}_t^{\mathrm{ml}}$} is \rev{recovered without the seed}.

ML-based recovery eliminates the shared-secret dependency by using an XGBoost model to predict expected measurements and a mixed-integer program (MIP) to match them against received data, yielding the lowest EDR at full deception and full resilience to seed compromise, but at roughly $60\times$ the latency of seed-based recovery. The hybrid strategy (60\% seed, 40\% ML) achieves the best overall balance: \rev{lower PSA than pure ML at four of the five deception levels (1.27\% vs.\ 1.96\% at full deception), within 0.7 percentage points of the uncompromised seed-only strategy at 60--100\% deception,} competitive CSER, and \rev{a $5\times$ latency reduction over pure ML (17.99 vs.\ 90.86\,ms at full deception)}, since the MIP solver operates on a reduced sensor subset, quadratically shrinking the search space. \rev{Because only the seed-routed 60\% depend on the key, a seed compromise cannot corrupt the ML-routed 40\%; quantifying the hybrid's end-to-end accuracy under compromise is left to future work.} Seed-based recovery is fastest but brittle, ML-based recovery is robust but computationally expensive, and the hybrid configuration delivers operational-grade latency with partial resilience, making it the recommended baseline for ROAM deployments.

\subsection{Performance of DQN-MTD in Hybrid Operating Mode}
\label{sb: hybrid_mode}
\vspace{-0.1cm}

To further test this flexibility, the RL agent \rev{flags the sensors to remap and chooses the seed ratio $\rho_t$ at every decision step, which assigns each flagged sensor to the seed-based or prediction-based pipeline} rather than applying a single global mode. The EMS routes each sensor's measurements through \rev{its assigned} pipeline, merges the individually remapped groups into a unified measurement vector, and forwards it to the SE-BDD module. This allows the agent to \rev{shift the recovery mix} based on system state, threat conditions, or seed availability, improving overall adaptability and resilience.

\begin{figure}[ht]
    \centering
        \subfigure[]{
            \label{fig:psa_118}
            \includegraphics[width=0.46\columnwidth]{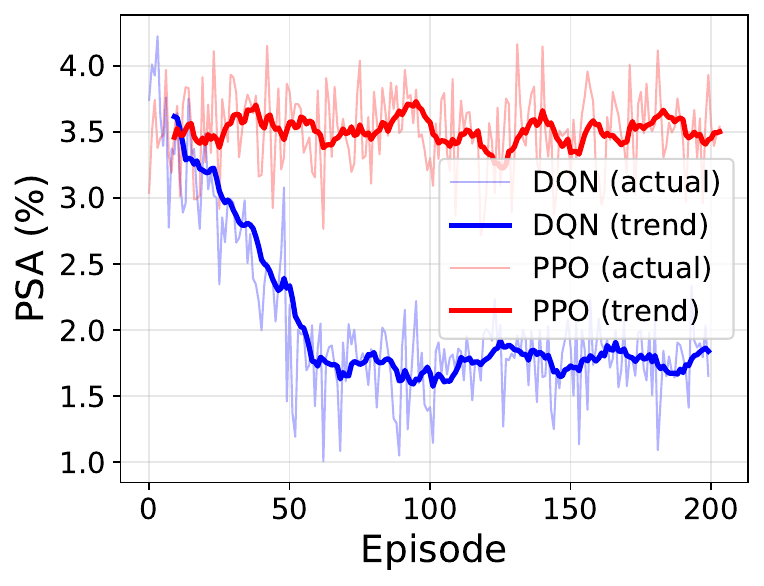}
        }
        \subfigure[]{
            \label{fig:pea_118}
            \includegraphics[width=0.45\columnwidth]{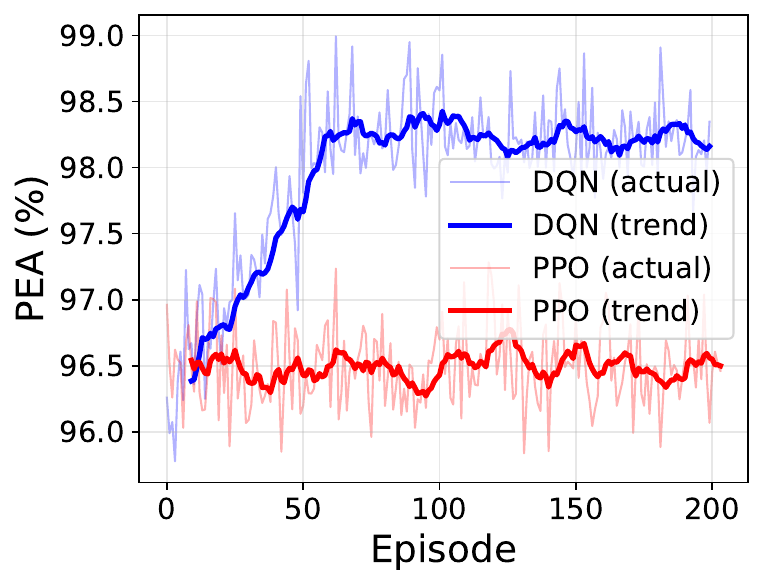}
        }
        \vspace{-6pt}
    \caption{ (a) Percentage of stealthy attack\rev{s} and (b) percentage of exposed attack\rev{s}.}
    \label{plot:psa_pea}
            \vspace{-8pt}
\end{figure}

\begin{figure}[ht]
    \centering
        \subfigure[]{
            \label{fig:edr_118}
            \includegraphics[width=0.46\columnwidth]{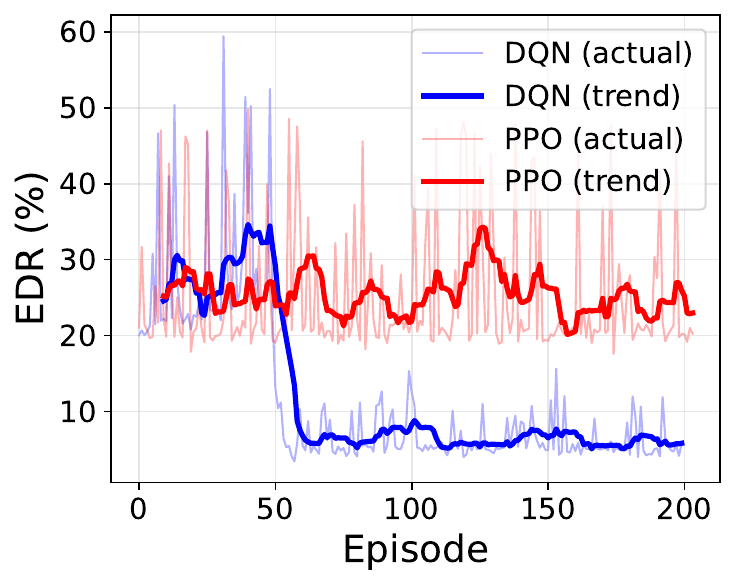}
        }
        \subfigure[]{
            \label{fig:cser_118}
            \includegraphics[width=0.45\columnwidth]{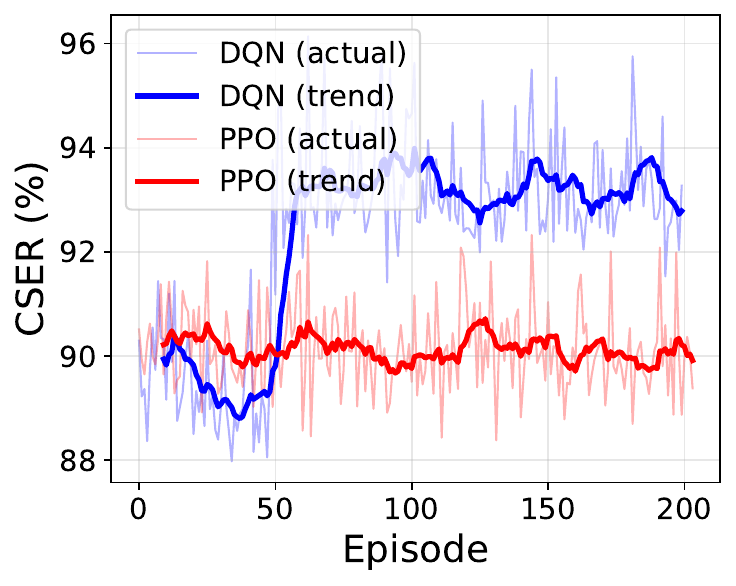}
        }
        \vspace{-6pt}
    \caption{ (a) \rev{Estimation deviation rate} and (b) compromised sensor elimination rate.}
    \label{plot:edr_cser}
            \vspace{-8pt}
\end{figure}

Fig.~\ref{action_hybrid} shows that after an initial exploration phase, the \rev{DQN} agent predominantly favors seed-based remapping\rev{, routing about 88\% of remapped sensors through the seed and the rest through prediction-based recovery}. The agent converges to shuffling roughly 75\% of sensors \rev{per step} (Fig.~\ref{deception_hybrid}), with cumulative reward (Fig.~\ref{rewards_hybrid}) and detection rate, which stabilizes above 90\% (Fig.~\ref{detection_hybrid}), converging accordingly. State deviation remains below 10\% (Fig.~\ref{deviations_hybrid}), and Fig.~\ref{attack_accuracy_hybrid} shows attack \rev{accuracy} falling below 30\%\rev{, and Fig.~\ref{fig:psa_118} shows PSA falling from about 3.5\% to 1.8\%, with PEA, EDR, and CSER in Figs.~\ref{plot:psa_pea}-\ref{plot:edr_cser},} confirming that the learned policy balances detection, system stability, and resource efficiency.The PPO baseline stays near the midpoints of the action ranges throughout training, which Section~\ref{sec:comparative_analysis} discusses.

\begin{table*}[!t]
\centering
\caption{Comparative Training Performance of DQN and PPO MTD Agents Across IEEE Bus Systems}
\label{tab:dqn_ppo_comparison}
\renewcommand{\arraystretch}{1.15}
\setlength{\tabcolsep}{4pt}
\begin{tabular}{p{0.2\columnwidth} p{0.30\columnwidth} p{0.15\columnwidth}p{0.15\columnwidth} p{0.15\columnwidth}p{0.15\columnwidth} p{0.15\columnwidth}p{0.15\columnwidth}}
\toprule
 & & \multicolumn{2}{c}{\textbf{IEEE 14-Bus}} & \multicolumn{2}{c}{\textbf{IEEE 57-Bus}} & \multicolumn{2}{c}{\textbf{IEEE 118-Bus}} \\
\cmidrule(lr){3-4} \cmidrule(lr){5-6} \cmidrule(lr){7-8}
\textbf{Category} & \textbf{Metric} & \textbf{DQN} & \textbf{PPO} & \textbf{DQN} & \textbf{PPO} & \textbf{DQN} & \textbf{PPO} \\
\midrule
\multirow{4}{*}{\rotatebox{90}{\scriptsize\textbf{System}}}
 & Buses / Lines        & \multicolumn{2}{c}{14 / 20}   & \multicolumn{2}{c}{57 / 80}   & \multicolumn{2}{c}{118 / 186} \\
 & Measurements ($|z|$) & \multicolumn{2}{c}{55}         & \multicolumn{2}{c}{218}        & \multicolumn{2}{c}{491}       \\
 & Scenarios (train/eval) & \multicolumn{2}{c}{1,000 / 500} & \multicolumn{2}{c}{1,000 / 500} & \multicolumn{2}{c}{1,000 / 500} \\
 & Training Episodes    & 200          & 204             & 200          & 204             & 200          & 204            \\
\midrule
\multirow{3}{*}{\rotatebox{90}{\scriptsize\textbf{Compute}}}
 & Training Time (h)    & \textbf{0.47}& 1.09            & \textbf{6.90}& 11.79           & \textbf{5.92}& 26.38          \\
 & Time/Episode (s)     & \textbf{8.34}& 10.82           & 123.24       & \textbf{85.95}  & \textbf{105.65}& 359.60        \\
 & Final Avg.\ Reward   & \textbf{$-$16,059}& $-$47,885  & \textbf{$-$47,784}& $-$78,717 & \textbf{$-$3,130}& $-$28,943   \\
\midrule
\multirow{3}{*}{\rotatebox{90}{\scriptsize\textbf{Policy}}}
 & Deception (\%)       & \textbf{73.84}{\tiny$\pm$0.44} & 52.85{\tiny$\pm$1.22} & \textbf{64.33}{\tiny$\pm$0.49} & 52.62{\tiny$\pm$1.23} & \textbf{73.92}{\tiny$\pm$0.41} & 52.84{\tiny$\pm$1.29} \\
 & Attack Acc.\ (\%)    & \textbf{32.85}{\tiny$\pm$1.36} & 50.04{\tiny$\pm$1.61} & \textbf{44.40}{\tiny$\pm$1.33} & 46.70{\tiny$\pm$1.68} & \textbf{26.11}{\tiny$\pm$1.39} & 35.25{\tiny$\pm$1.63} \\
 & Seed ratio $\rho$ (\%)  & \textbf{87.84}{\tiny$\pm$0.87} & 50.25{\tiny$\pm$1.85} & \textbf{69.07}{\tiny$\pm$0.57} & 50.24{\tiny$\pm$2.00} & \textbf{88.13}{\tiny$\pm$0.62} & 50.40{\tiny$\pm$1.93} \\
\midrule
\multirow{4}{*}{\rotatebox{90}{\scriptsize\textbf{Defense}}}
 & PSA (\%)  $\downarrow$  & \textbf{2.33}{\tiny$\pm$0.22}  & 4.58{\tiny$\pm$0.35}  & \textbf{3.30}{\tiny$\pm$0.28}  & 3.72{\tiny$\pm$0.34}  & \textbf{1.78}{\tiny$\pm$0.23}  & 3.48{\tiny$\pm$0.30}  \\
 & PEA (\%)  $\uparrow$   & \textbf{97.67}{\tiny$\pm$0.22} & 95.42{\tiny$\pm$0.35} & \textbf{96.70}{\tiny$\pm$0.28} & 96.28{\tiny$\pm$0.34} & \textbf{98.22}{\tiny$\pm$0.23} & 96.52{\tiny$\pm$0.30} \\
 & CSER (\%) $\uparrow$   & \textbf{92.90}{\tiny$\pm$0.57} & 90.85{\tiny$\pm$0.71} & \textbf{92.57}{\tiny$\pm$0.61} & 92.04{\tiny$\pm$0.70} & \textbf{93.18}{\tiny$\pm$0.87} & 90.11{\tiny$\pm$0.82} \\
 & EDR (\%)  $\downarrow$  & \textbf{14.88}{\tiny$\pm$2.32} & 23.15{\tiny$\pm$1.93} & \textbf{35.17}{\tiny$\pm$3.43} & 54.17{\tiny$\pm$4.63} & \textbf{6.14}{\tiny$\pm$2.33}  & 25.46{\tiny$\pm$9.08} \\
\bottomrule
\end{tabular}
\vspace{2pt}
\begin{flushleft}
\footnotesize
\textit{Note:} All values report mean $\pm$ std over the last 100 training episodes for both algorithms.
Bold values indicate the better-performing algorithm per bus system.
$\downarrow$\,=\,lower is better; $\uparrow$\,=\,higher is better.
\end{flushleft}

\vspace{-12pt}
\end{table*}

\subsection{Comparative Analysis of DQN and PPO MTD Agents}
\label{sec:comparative_analysis}

Table~\ref{tab:dqn_ppo_comparison} compares DQN and PPO trained on the identical MTD task across the IEEE 14-, 57-, and 118-bus systems, averaged over the last 100 training episodes. DQN trains substantially faster at every scale, with the gap reaching $4.5\times$ on the 118-bus system, since it samples mini-batches from a replay buffer, whereas PPO collects full on-policy rollouts whose cost grows with system complexity.

The two algorithms converge to qualitatively different defense postures: DQN consistently selects high deception coverage \rev{(64-74\%)} and routes most remapped sensors through seed-based recovery \rev{($\rho \approx 69$--$88\%$)}, while PPO \rev{averages about 53\% deception and a 50\% seed ratio on every topology}. These values sit at the midpoints of the action ranges, indicating that PPO's stochastic training-time policy had not concentrated on a preferred configuration within the training budget; its deterministic post-training policy (Section~\ref{sec:post_training}) is markedly different.

This policy divergence translates directly into performance: DQN achieves lower \rev{PSA}, higher detection, and better state estimation across all systems \rev{during training}, most strikingly on the 118-bus system, where its high-seed-recovery strategy yields a $4.1\times$ \rev{lower EDR than} PPO's balanced split. The \rev{PSA and detection gaps narrow} on the 57-bus system, consistent with DQN's comparatively lower deception there.

\begin{table*}[!t]
\centering
\caption{Post-Training Evaluation of DQN and PPO MTD Agents Across IEEE Bus Systems}
\label{tab:eval_comparison}
\renewcommand{\arraystretch}{1.15}
\setlength{\tabcolsep}{4pt}
\begin{tabular}{@{}l l cc cc cc@{}}
\toprule
 & & \multicolumn{2}{c}{\textbf{IEEE 14-Bus}} & \multicolumn{2}{c}{\textbf{IEEE 57-Bus}} & \multicolumn{2}{c}{\textbf{IEEE 118-Bus}} \\
\cmidrule(lr){3-4} \cmidrule(lr){5-6} \cmidrule(lr){7-8}
\textbf{Category} & \textbf{Metric} & \textbf{DQN} & \textbf{PPO} & \textbf{DQN} & \textbf{PPO} & \textbf{DQN} & \textbf{PPO} \\
\midrule
\multirow{2}{*}{\rotatebox{90}{\scriptsize\textbf{Policy}}}
 & Deception (\%)       & 75.00{\tiny$\pm$0.00} & 50.00{\tiny$\pm$0.02} & 70.00{\tiny$\pm$0.00} & 79.71{\tiny$\pm$0.69} & 75.00{\tiny$\pm$0.00} & \rev{80.00}{\tiny$\pm$0.00} \\
 & Attack Acc.\ (\%)    & \textbf{31.55}{\tiny$\pm$1.46} & 41.85{\tiny$\pm$1.58} & \textbf{45.34}{\tiny$\pm$1.55} & 49.96{\tiny$\pm$1.13} & 25.59{\tiny$\pm$1.30} & \textbf{13.13}{\tiny$\pm$0.96} \\
\midrule
\multirow{4}{*}{\rotatebox{90}{\scriptsize\textbf{Defense}}}
 & PSA (\%)  $\downarrow$  & \textbf{2.16}{\tiny$\pm$0.18}  & 4.86{\tiny$\pm$0.37}  & \textbf{2.91}{\tiny$\pm$0.26}  & 3.17{\tiny$\pm$0.26}  & 1.70{\tiny$\pm$0.25}  & \textbf{1.05}{\tiny$\pm$0.15}  \\
 & PEA (\%)  $\uparrow$   & \textbf{97.84}{\tiny$\pm$0.18} & 95.14{\tiny$\pm$0.37} & \textbf{97.09}{\tiny$\pm$0.26} & 96.83{\tiny$\pm$0.26} & 98.30{\tiny$\pm$0.25} & \textbf{98.95}{\tiny$\pm$0.15} \\
 & CSER (\%) $\uparrow$   & \textbf{93.15}{\tiny$\pm$0.45} & 88.39{\tiny$\pm$0.78} & 93.57{\tiny$\pm$0.59} & \textbf{93.65}{\tiny$\pm$0.58} & \textbf{93.38}{\tiny$\pm$0.86} & 92.02{\tiny$\pm$1.00} \\
 & EDR (\%)  $\downarrow$  & \textbf{14.64}{\tiny$\pm$2.31} & 30.09{\tiny$\pm$1.70} & 31.88{\tiny$\pm$3.66} & \textbf{19.91}{\tiny$\pm$2.10} & \textbf{5.01}{\tiny$\pm$2.01}  & 17.65{\tiny$\pm$8.48} \\
\midrule
 & Reward               & \textbf{$-$13,858}{\tiny$\pm$2,302} & $-$40,366{\tiny$\pm$2,888} & $-$48,970{\tiny$\pm$2,839} & \textbf{$-$44,395}{\tiny$\pm$1,550} & $-$1,957{\tiny$\pm$1,986} & \textbf{9,670}{\tiny$\pm$6,852} \\
\bottomrule
\end{tabular}
\vspace{2pt}
\begin{flushleft}
\footnotesize
\textit{Note:} All values report mean $\pm$ std over 50 post-training evaluation episodes.
Bold values indicate the better-performing algorithm per bus system.
$\downarrow$\,=\,lower is better; $\uparrow$\,=\,higher is better.
\end{flushleft}
\vspace{-10pt}
\end{table*}

\begin{table*}[!t]
\centering
\caption{Adversarial Evaluation of DQN and PPO MTD Agents Under Adaptive SAC Attacker}
\label{tab:adversarial_eval}
\renewcommand{\arraystretch}{1.15}
\setlength{\tabcolsep}{4pt}
\begin{tabular}{@{}l l cc cc cc@{}}
\toprule
 & & \multicolumn{2}{c}{\textbf{IEEE 14-Bus}} & \multicolumn{2}{c}{\textbf{IEEE 57-Bus}} & \multicolumn{2}{c}{\textbf{IEEE 118-Bus}} \\
\cmidrule(lr){3-4} \cmidrule(lr){5-6} \cmidrule(lr){7-8}
\textbf{Category} & \textbf{Metric} & \textbf{DQN} & \textbf{PPO} & \textbf{DQN} & \textbf{PPO} & \textbf{DQN} & \textbf{PPO} \\
\midrule
\multirow{4}{*}{\rotatebox{90}{\scriptsize\textbf{Defense}}}
 & PSA (\%)  $\downarrow$  & \textbf{2.57}{\tiny$\pm$0.43}  & 4.26{\tiny$\pm$0.53}  & \textbf{1.78}{\tiny$\pm$0.51}  & 2.60{\tiny$\pm$0.32}  & \textbf{1.14}{\tiny$\pm$0.32}  & 1.15{\tiny$\pm$0.29}  \\
 & PEA (\%)  $\uparrow$   & \textbf{97.43}{\tiny$\pm$0.43} & 95.74{\tiny$\pm$0.53} & \textbf{98.22}{\tiny$\pm$0.51} & 97.40{\tiny$\pm$0.32} & \textbf{98.86}{\tiny$\pm$0.32} & \rev{98.85}{\tiny$\pm$0.29} \\
 & CSER (\%) $\uparrow$   & \textbf{92.46}{\tiny$\pm$0.89} & 89.04{\tiny$\pm$1.07} & 93.77{\tiny$\pm$0.77} & \textbf{95.25}{\tiny$\pm$0.48} & \textbf{92.13}{\tiny$\pm$1.70} & \rev{92.12}{\tiny$\pm$1.43} \\
 & EDR (\%)  $\downarrow$  & \textbf{38.21}{\tiny$\pm$8.13} & 38.82{\tiny$\pm$3.58} & 57.79{\tiny$\pm$5.13} & \textbf{38.54}{\tiny$\pm$3.91} & 20.56{\tiny$\pm$10.83} & \textbf{19.39}{\tiny$\pm$10.32} \\
\midrule
\multirow{2}{*}{\rotatebox{90}{\scriptsize\textbf{Attack}}}
 & Attack Acc.\ (\%) $\downarrow$ & \textbf{33.97}{\tiny$\pm$2.22} & 38.83{\tiny$\pm$1.32} & \textbf{29.42}{\tiny$\pm$11.71} & 54.59{\tiny$\pm$2.21} & \textbf{14.23}{\tiny$\pm$1.46} & 14.39{\tiny$\pm$1.54} \\
 & Reward                        & \textbf{$-$37,134}{\tiny$\pm$8,909}     & $-$45,347{\tiny$\pm$2,732}     & \textbf{$-$59,629}{\tiny$\pm$10,209}    & $-$66,885{\tiny$\pm$6,417}     & 5,418{\tiny$\pm$7,235}      & \textbf{6,160}{\tiny$\pm$7,064}      \\
\bottomrule
\end{tabular}
\vspace{2pt}
\begin{flushleft}
\footnotesize
\textit{Note:} All values report the mean $\pm$ std over 50 adversarial evaluation episodes with the fully trained SAC attacker.
Bold values indicate the better-performing algorithm per bus system.
$\downarrow$\,=\,lower is better; $\uparrow$\,=\,higher is better.
\end{flushleft}
\vspace{-12pt}
\end{table*}

\subsection{Post-Training Evaluation of DQN and PPO MTD Agents}
\label{sec:post_training}

Both DQN and PPO were evaluated over 50 held-out episodes with fixed, deterministic policies (Table~\ref{tab:eval_comparison}). The two algorithms deploy notably different levels of deception: DQN maintains a consistent 70-75\% across all topologies, while PPO is conservative on the 14-bus system (50\%) but escalates to 80\% on the larger systems, shaping the performance trade-offs below. Neither algorithm uniformly dominates. DQN achieves higher CSER and markedly better EDR on the 14-bus and 118-bus systems, where its high-deception, seed-dominant policy produces superior recovery\rev{, and it also achieves lower PSA on the 14- and 57-bus systems}. 

In the 57-bus system, PPO's higher deception reverses the EDR comparison by a factor of 1.6 and yields a higher reward. The most striking reversal occurs on the 118-bus system, where PPO's 80\% deception achieves the lowest overall PSA (1.05\% \rev{vs.\ 1.70\%} for DQN) \rev{and attack accuracy (13.13\% vs.\ 25.59\%)}, though DQN retains the advantage in CSER and EDR. On the 14-bus system, DQN outperforms PPO across all metrics because PPO's conservative 50\% deception provides the BDD module with insufficient discriminative evidence. These results point to complementary algorithmic strengths: DQN provides more consistent detection and state recovery on small- and large-scale topologies, while PPO \rev{deploys more aggressive deception on the 57- and 118-bus systems}. No single algorithm uniformly dominates, suggesting that topology-aware agent selection, or an ensemble strategy, may yield the strongest overall defense posture.

\subsection{Evaluation with Adversarial Multi-Agent Framework}
\label{sec:adversarial}

Table~\ref{tab:adversarial_eval} evaluates both defenders against a fully-trained SAC attacker over 50 deterministic episodes. Relative to Table~\ref{tab:eval_comparison}, EDR degrades by $2.6\times$ for DQN \rev{(14-bus)} and $1.9\times$ for PPO \rev{(57-bus)} on their respective most vulnerable topologies, confirming that the SAC attacker exploits policy-specific weaknesses. PSA and CSER standings from standard evaluation largely hold: DQN achieves lower stealthy-attack success on the 14-bus and 57-bus systems, while on the 118-bus system, both converge to near-identical PSAs, suggesting the large sensor space dilutes targeted attacks regardless of strategy. \rev{On the 57-bus system, PPO's small CSER lead from standard evaluation widens (95.25\% vs.\ 93.77\%), as its higher deception gives the BDD module richer discriminating features.}

The most consequential finding concerns EDR on the 57-bus topology: DQN's adversarial EDR reaches 57.79\%, nearly double its standard-evaluation value, while PPO's rises only to 38.54\%. This exposes a structural vulnerability. DQN displaces more attacks in this topology, yet suffers from worse state recovery, indicating that displacement alone is insufficient when the recovery mechanism cannot reconstruct corrupted estimates against an adaptive adversary. On the 118-bus system, both algorithms converge to comparable EDR with positive mean rewards, confirming effective defense at scale. Overall, PPO shows greater \rev{state-recovery} robustness under adversarial pressure with tighter \rev{EDR} variance, while DQN retains \rev{lower PSA on every topology and stronger displacement on the 14- and 57-bus systems}, reinforcing the case for an ensemble or topology-aware deployment strategy.

\begin{figure}[htbp]
    \centering
    \vspace{-2pt}
    \includegraphics[width=0.90\columnwidth]{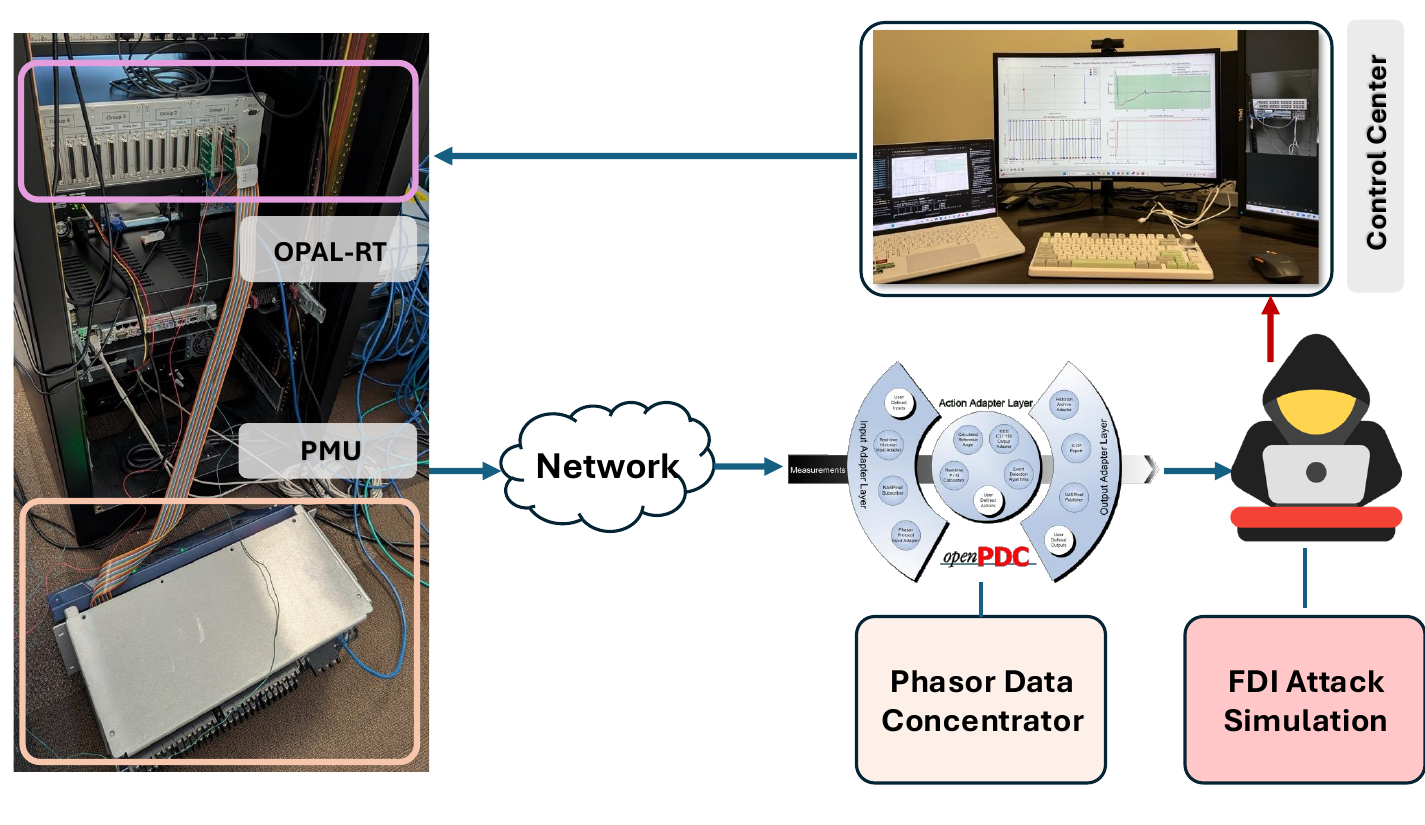}
    \vspace{-2pt}
    \caption{Implementation of ROAM \rev{on the} real-time HIL testbed.}
    \label{fig:test_bed}
    \vspace{-10pt}
\end{figure}

\begin{table}[h]
\centering
\caption{Real-Time ROAM Performance Summary}
\label{tab:results}
\begin{tabular}{lc}
\hline
\textbf{Metric} & \textbf{Value} \\
\hline
Detection rate \rev{(end of session)} & 83.2\% \\
\rev{Attacked frames nullified by MTD} & 290 / 342 \rev{(84.8\%)} \\
Shuffle interval (configured) & 3{,}000 ms \\
Avg.\ inter-shuffle gap (measured) & 2{,}995.3 ms \\
Avg.\ shuffle execution time & 17.97 ms \\
Max.\ shuffle execution time & 56.28 ms \\
Avg.\ AGC reshuffle latency & 77.86 ms \\
Max.\ AGC reshuffle latency & 382.77 ms \\
\rev{Blind window fraction (avg.\ / max.)} & \rev{2.6\% / 12.8\%} \\
\hline
\end{tabular}
\vspace{-10pt}
\end{table}
\subsection{Real-Time System Integration}
\label{sec: real_time} 



To validate the proposed framework beyond simulation, we deployed the trained DQN on a live OPAL-RT HIL testbed (Fig.~\ref{fig:test_bed}), replicating the IEEE 118-bus system with 18 virtual \rev{PMUs} and 1 hardware PMU connected to a custom Python-based PDC over IEEE C37.118. The real-time layer adds the complexity of physical device scheduling, OS jitter, and live TCP/IP streams beyond the 118-bus simulation used for training; this PDC orchestrates the MTD framework by receiving real-time PMU data and shuffling IDs based on the DQN agent's suggested deception amount. The DQN inference engine runs as an isolated subprocess, producing MTD decisions every 3 seconds and writing them to a shared JSON file; the PDC acquisition process reads these decisions and applies seed-based, ML-based, or hybrid PMU ID remapping at each shuffle interval, while a feedback daemon stores detection rate and attacked PMU IDs back to the deployer for closed-loop adaptation. 

A coordinated FDI attack targeting multiple PMUs simultaneously was injected throughout the session; Table~\ref{tab:results} summarizes the outcomes. The \rev{running} detection rate improved monotonically from 0\% to \textbf{83.2\%} \rev{by the end of the session, and 290 of the 342 attacked frames (84.8\%) were nullified by the MTD}, confirming effective sim-to-real transfer Shuffle execution averaged 17.97~ms for 19 PMUs \rev{(56.28~ms maximum)}, and AGC reshuffle latency (mapping commit to first frame processed under the new assignment) averaged 77.86~ms, \rev{an average blind window of 2.6\% of the 3-second shuffle cycle; the worst-case latency of 382.77~ms corresponds to 12.8\%, still well below one cycle.}

\section{Conclusions}
\label{sec:conclusion}

This paper presented ROAM, a DRL-orchestrated hybrid moving-target defense that continuously randomizes the sensor-to-ID mapping in wide-area measurement systems. By learning, at each decision epoch, how many sensors to remap\rev{, which sensors to flag, and how to split their recovery between the seed-based and ML-based pathways}, ROAM \rev{keeps the percentage of stealthy attacks below 3\% and sustains detection above 93\% on held-out scenarios} across the IEEE 14-, 57-, and 118-bus systems, while the hybrid pathway \rev{confines the impact of a seed compromise to the seed-routed subset}. Adversarial co-evolution against a learning SAC attacker \rev{kept stealthy attacks below 2.6\% but raised state-estimation deviation by up to $2.6\times$}, revealing topology-dependent trade-offs between the value-based and policy-gradient defenders \rev{that} motivate an ensemble or topology-aware deployment. Real-time OPAL-RT experiments demonstrated direct sim-to-real transfer within the timing constraints of operational WAMPAC systems\rev{, with a worst-case reshuffle blind window of 12.8\% of the decision cycle}. Future work will \rev{evaluate the hybrid pathway under seed compromise end to end and} pursue federated training for multi-utility deployment without sharing topology data, extension to DER-rich distribution networks, and formal privacy guarantees for the prediction module.

\section*{Acknowledgement}
This work is supported by the Department of Energy (DOE) under Award\# DE-CR0000024. Any opinions, findings, conclusions, or recommendations expressed in this material are those of the authors and do not necessarily reflect the
views of the DOE.
\vspace{-0.2cm}

\section*{Reproducibility and Code Availability}
To support reproducibility, trained DQN/PPO agent weights, attack scenario datasets (1{,}000 training~+~500 evaluation BDD-compliant FDI scenarios), and evaluation scripts are publicly available at ~\cite{roam_drl_mtd}. 



\bibliographystyle{ieeetr}
\bibliography{sections/References}

\end{document}